\documentclass[twocolumn, prd, a4paper, superscriptaddress, nofootinbib, showpacs]{revtex4}

\usepackage{graphicx}
\usepackage{amsfonts}
\usepackage{amsmath,amssymb}

\newcommand{\lb}[0] { \left( }
\newcommand{\rb}[0] { \right) }
\newcommand{\beqs} { \begin{eqnarray} }
\newcommand{\eeqs} { \end{eqnarray} }
\newcommand{\bsub} { \begin{subequations} }
\newcommand{\esub} { \end{subequations} }
\newcommand{\nn} {\nonumber}
\newcommand{\ep}[0] { \epsilon }
\renewcommand{\eqref}[1]{(\ref{#1})}
\newcommand{\EE}[2]{#1 \times 10^{#2}}

\begin{document}

\title{Parameterization of the energy and rapidity distributions of 
secondary pions and kaons produced in energetic proton -- proton collisions}

\date{\today}

\author{Hylke~B.~J.~Koers}
\email{hkoers@nikhef.nl}
\affiliation{NIKHEF, P.O. Box 41882, 1009 DB Amsterdam, The Netherlands}
\affiliation{University of Amsterdam, Amsterdam, The Netherlands}
\author{Asaf~Pe'er}
\author{Ralph~A.~M.~J.~Wijers}
\affiliation{Astronomical Institute `Anton Pannekoek', Faculty of Science, University of Amsterdam, Kruislaan 403, 1098 SJ Amsterdam, The Netherlands}

\begin{abstract} 
The possibility of proton acceleration to very high energies in
astrophysical sources may have unique observational consequences.
In particular, the decay of secondary mesons created in the interaction of
energetic protons with photons or nucleons gives rise to high-energy gamma rays
and neutrinos with potentially observable fluxes.
Recently, there has been considerable interest in the parameterization of the
energy spectra of these secondaries. Less attention has been paid to the
angular distributions, which may have an important effect on observational
quantities and are required to address collisions between protons with different energies
and an arbitrary scattering angle.
In this work, we  study the complete particle distributions
of secondary mesons created in proton -- proton collisions.
We present parameterizations
of the energy and rapidity distributions of secondary pions and kaons that
reproduce results generated with the event generator 
\texttt{PYTHIA} to within $\sim$10\% 
in the bulk of the parameter space.
The parameterizations are based on incident proton energies from 1 TeV to 1 PeV
and are suited for extrapolation to higher energies.
We present several applications of these parameterizations.
Energy spectra and angular distributions of stable decay products
(electrons, neutrinos and gamma rays) follow readily.
We give an example of the gamma-ray spectrum that results from the decay
of $\pi^0$ mesons created in a proton -- proton collision.
We show that there is a strong correlation between the energy of
secondary mesons and the
degree of collimation around the direction of the colliding protons.
This effect  may have important implications for the detection possibility
of neutrinos created in the interaction of a developing GRB with its
surroundings.
\end{abstract}

\preprint{NIKHEF-2006-010}

\maketitle

\section{Introduction}

The possibility of proton acceleration to very high energies in
astrophysical sources may provide unique observational  opportunities.
The interaction of energetic protons with photons or nucleons results in
copious production of secondary mesons decaying into high-energy
gamma rays and neutrinos that can be observed with current and future
detectors.
The recently observed TeV gamma-ray emission from supernova remnant 
\mbox{RX J1713.7-3946} \cite{Aharonian:2005qm}
has been attributed to this mechanism \cite{Butt:2001ff},
although such an origin is still under debate \cite{Reimer:2002ea}.
TeV gamma rays have been reported in coincidence with gamma-ray burst (GRB)
970417a \cite{2000ApJ...533L.119A, 2003ApJ...583..824A} but also in this case
it is not established whether the origin is hadronic \cite{2005ApJ...633.1018P}.

The existence of astrophysical proton accelerators is indicated by observations
of high-energy cosmic rays (CRs).
There is evidence for a substantial proton component 
above the `knee' at  \mbox{$\sim$$\EE{4}{6}$} GeV in the cosmic-ray spectrum
(see e.g. Ref. \cite{Bhattacharjee:1998qc} for a review).
Observations of extensive air showers
due to CRs with energies up to \mbox{$\sim$$10^{11}$} GeV are consistent with 
nucleon primaries, although other primaries are also possible
(e.g., Ref. \cite{Halzen:1994gy}).

Various astrophysical systems have been suggested as CR sources.
Galactic supernova remnants are the leading candidate for the generation of
CRs with energies up to \mbox{$\sim$$10^{8}$}~GeV
(see e.g. Ref. \cite{Biermann:1995qy}). Several extragalactic sources have been considered
as possible sources of higher energy CRs, such as
active galactic nuclei \cite{Berezinsky:2002vt} (see however Ref. \cite{1995ApJ...454...60N}),
hot spots of Fanaroff-Riley class II radio galaxies
\cite{1993A&A...272..161R, 1995ApJ...454...60N},
pulsars \cite{Venkatesan:1996jw}
and GRBs \cite{Vietri:1995hs, Waxman:1995vg}.

A population of high-energy protons in these sources 
would carry a rich phenomenology. In GRBs for example, the interaction of
accelerated protons with GRB photons leads to  
$\sim10^5$ GeV neutrinos \cite{Waxman:1997ti} and to $\sim$$10^2 - 10^3$
GeV gamma rays \cite{1998ApJ...499L.131B,2005ApJ...633.1018P}.
High-energy proton interactions may play an important role in the interaction of a
developing GRB with its environment, e.g. when the 
fireball has not yet emerged from 
the stellar surface \cite{Meszaros:2001ms, Razzaque:2003uv}
or when energetic GRB protons collide with cold protons in
the GRB surroundings \cite{Granot:2002qz,Razzaque:2002kb,Razzaque:2003uw}.

Detailed parameterizations of the energy spectra of 
secondary particles created in proton -- proton ($pp$) collisions
are essential in the study of 
particle production in astrophysical proton accelerators.
Such parameterizations were recently presented by 
\citet{Kamae:2006bf} and \citet{Kelner:2006tc}.
However, the parameterizations presented by these authors
do not include the
angular distributions of the secondary particles.
As a consequence,
these parameterizations can only be applied to the scattering geometry for which they
were derived, viz. with a target proton at rest.
This is sufficient if one of the protons
is at rest in the observer frame, but a more general treatment is
required if the bulk of the protons can be accelerated.

The parameterization of the complete particle distributions
is an important generalization
because this provides, through
Lorentz transformations, 
secondary particle distributions and energy spectra after a collision of two protons
with arbitrary energies and an arbitrary incident angle.
Such a parameterization can therefore be applied to any acceleration scenario.
A second advantage is that the full distribution includes
correlations between the energy and the angle of outgoing particles,
which may have important effects on observable quantities in a non-isotropic
environment.

\citet{Badhwar:1977zf}, \citet{1981Ap&SS..76..213S} and  \citet{Blattnig:2000zf} have 
presented parameterizations of the complete distributions of charged and neutral pions and charged kaons created in $pp$ collisions.
However, these parameterizations are valid for incident proton energies $E_p \lesssim \EE{2}{3}$ GeV,
which is much lower than the highest proton energies $\sim$$10^{11}$ GeV expected
in accelerating astrophysical sources.

In this paper we study the complete distributions of
secondary particles produced in $pp$ collisions
through Monte Carlo simulations. 
 We consider
a proton with energy \mbox{$10^3$ GeV $< E_p <  10^6$ GeV} that collides with
a proton at rest, which corresponds to center-of-mass energy 
\mbox{$  43 \textrm{ GeV} < \sqrt{s} < \EE{1.4}{3} \textrm{ GeV} $}. We assume that the
distribution of a secondary particle species is invariant under rotations around the collision axis, which
implies that the distribution is fully parameterized with two independent kinematical
variables. We present parameterizations of the energy and rapidity distributions 
of secondary pions and kaons. The
parameterizations are based on Monte Carlo
data in the simulated energy range but can be applied to collisions of protons
with higher energies.

We consider only secondary pions and kaons
and not their stable decay products, viz. electrons, neutrinos and gamma rays.
This approach separates the physics in the $pp$ collision
from subsequent decay processes. Energy spectra and particle distributions of the resulting 
stable daughter particles are readily found from our results (either analytically or as part of
a computer code) and the
well-known decay spectra of pions and kaons (see e.g. Refs. \cite{HalzenMartin,Kelner:2006tc}).
We do not separately consider short-lived mesons (such as $\eta, \rho$ or $\omega$)
because their lifetime is much shorter than that of charged kaons and
pions. The decay products of these mesons, mostly pions and gamma rays, are 
grouped together with the prompt secondaries.
The restriction to the $pp$ interaction \emph{per se} gives our results a
broad applicability.
For example, it has been pointed out recently \cite{Kashti:2005qa,Asano:2006zz} that energy losses
of pions and kaons can leave an imprint on the energy spectra of the daughter particles
in GRB jets. A proper treatment of this effect requires knowledge of the pion
and kaon distributions.

Hadron interactions have a complex phenomenology due to the compositeness of the ingoing and
outgoing particles.
It is currently not possible to compute the cross section or the resulting particle distribution from
first principles. 
Therefore, detailed studies of particle production in hadron interactions require the use of 
event generators.
In this work, we simulate the  $pp$  interaction with the 
event generator \texttt{PYTHIA} \cite{Sjostrand:2003wg}, which
is tested against experimental data. It is capable of simulating various incident and target
particles so that it is
 possible to extend this work to
proton -- neutron and proton -- photon interactions with essentially the same
code. \texttt{PYTHIA} uses the `Lund string model' \cite{Andersson:1983ia} to 
describe the process of secondary meson formation.

This paper is organized as follows: in section \ref{sec:sigma_mult}, we present
 experimental data on the $pp$ cross section and the charged multiplicity,
i.e. the number of charged particles created in a single inelastic collision.
In section \ref{sec:cross_kin} we discuss the kinematics of the simulated interaction 
and introduce the particle distribution with respect to energy and rapidity.
Details on the event simulation
with \texttt{PYTHIA} and the fitting procedure are discussed
in section \ref{sec:method}.
In section \ref{sec:results}, we present a comparison between \texttt{PYTHIA} results and
experimental data and we present the parameterizations of the energy spectra and
particle distributions of secondary pions and kaons.
Applications  of these parameterizations are considered in section 
\ref{sec:app}. We demonstrate through explicit examples how the
parameterizations can be used to study particle production in collisions of protons with
different energies and an arbitrary incident angle. We also present an example
in which we derive the gamma-ray energy spectrum resulting from $\pi^0 \to \gamma \gamma$ decay.
In section \ref{sect:disc}, we discuss the application of the parameterizations
to incident protons with very high energies.
We discuss the results in section \ref{sec:discussion}.
Conclusions are presented in section \ref{sec:conclusion}.

\section{Experimental data on the cross section and secondary multiplicity
in proton -- proton interactions}
\label{sec:sigma_mult}
\begin{figure}
\begin{center}
\includegraphics[angle=270, width=8.6cm]{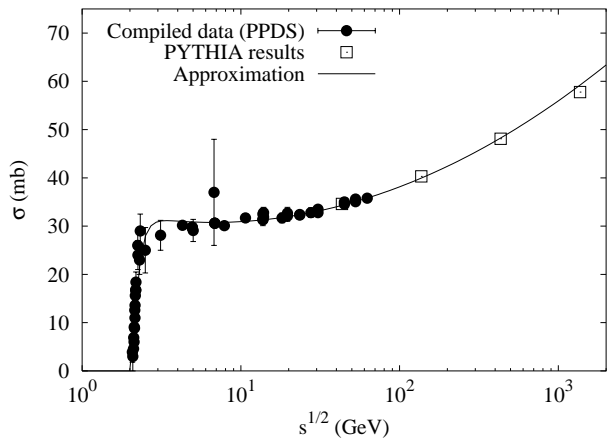}
\end{center}
\caption{Comparison between the inelastic $pp$ cross section calculated with
\texttt{PYTHIA} (open squares) and experimental data (disks). The solid line represents
the fit given in eq. \eqref{eq:sigmatotinel}.
Experimental data is taken from the PPDS (see footnote \ref{footnote:PPDS}).}
\label{fig:sigma}
\end{figure}

\begin{figure}
\begin{center}
\includegraphics[angle=270, width=8.6cm]{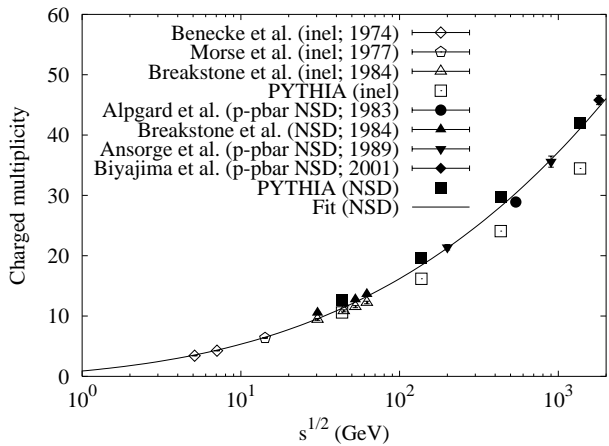}
\end{center}
\caption{Comparison between the charged multiplicity calculated 
with \texttt{PYTHIA} (squares) and experimental data (other symbols). 
Open symbols correspond to inelastic processes (inel), solid
symbols correspond to non-single-diffractive (NSD) processes.
The solid line shows the approximation given in eq. \eqref{eq:theory:chmult}.}
\label{fig:mult}
\end{figure}

In this section, we discuss experimental data on
the cross section of $pp$ interactions and on
the number of charged particles created in a $pp$ interaction.
The data presented in this section are used in section \ref{sec:results}
to validate our numerical method.

\subsection{Cross section}
Proton -- proton interactions are usually separated into elastic scattering, in which no particles are
created; diffractive interactions, in which the energy transfer between the protons is
small; and inelastic non-diffractive interactions, in which the energy transfer is large enough
for the constituent quarks and gluons to interact.
The total $pp$ cross section $\sigma_{\textrm{tot}}$ can be expanded in terms of
these processes as 
\beqs
\sigma_{\textrm{tot}} = \sigma_{\textrm{nd}}  + \sigma_{\textrm{sd}}
+ \sigma_{\textrm{dd}} + \sigma_{\textrm{el}} \, ,
\eeqs
where $\sigma_{\textrm{nd}}$, $\sigma_{\textrm{sd}}$, $\sigma_{\textrm{dd}}$ and $\sigma_{\textrm{el}}$
are the cross section for
non-diffractive processes (the hard QCD processes), single diffraction ($AB \to XB$ or $AB \to AX$),
double diffraction ($AB \to XY$), and elastic scattering ($AB \to AB$) respectively.
In this work we do not explicitly separate diffractive and non-diffractive processes
because we are mostly interested in astrophysical applications where it will be
impossible to distinguish between these components;
see Ref. \cite{Kamae:2006bf} for a separate treatment.

We are primarily interested in the inelastic  $pp$ cross section 
$\sigma_{\textrm{inel}} = \sigma_{\textrm{nd}} + \sigma_{\textrm{sd}}
+ \sigma_{\textrm{dd}}$
because all processes that create secondary particles are contained
in this quantity.
At energies above the threshold energy $E_{\textrm{th}}= 1.22$ GeV and below
$\sqrt{s} = \EE{3}{4} \textrm{ GeV}$, the inelastic
cross section for a proton with energy $E_p$ interacting with a target proton at rest
can be fitted with \cite{Kelner:2006tc}:
\beqs
\nn \sigma_{\textrm{inel}}^{\textrm{fit}} (E_p)  \hspace{-0.1cm} & = &  \hspace{-0.1cm}  \lb 33.24 - 3.624 \log E_p  + 1.325 \lb \log  E_p  \rb^2 \rb \\ 
\label{eq:fitsigmainel} & &  \hspace{-0.1cm}  \times  \, \lb 1- \lb \frac{E_{\textrm{th}}}{E_p} \rb^4 \rb^2 \textrm{ mb} \, ,
\eeqs
where $E_p$ is measured in GeV. In deriving this formula, it is assumed that 
the ratio of the inelastic cross section to the total cross section,
which for energies \mbox{$43$ GeV $ < \sqrt{s} < 63$ GeV} is given by \cite{Amaldi:1979kd}
\beqs
\label{eq:sigmatotinel}
\sigma_{\textrm{tot}} = 1.21 \,  \sigma_{\textrm{inel}} \, ,
\eeqs
holds for higher energies \mbox{63 GeV $ < \sqrt{s} <  \EE{3}{4} $ GeV} as well.
The incident proton energy $E_p$ in eq. \eqref{eq:fitsigmainel} is related to the center-of-mass
energy $\sqrt{s}$ as
\beqs
\label{eq:Ep:sqrts}
E_p = \frac{s}{2 m_p c^2} - m_p c^2 \, ,
\eeqs 
where $m_p$ is the proton mass.
In figure \ref{fig:sigma}, we show
the approximation given in eq. \eqref{eq:fitsigmainel} together with \texttt{PYTHIA} results
(see section \ref{sec:results} below) 
and the available experimental data.\footnote{\label{footnote:PPDS}A compilation of
experimental data on the inelastic $pp$ cross section is available 
at the Particle Physics Data System (PPDS) website
\texttt{http://wwwppds.ihep.su:8001/}. In producing figure \ref{fig:sigma}, we have not considered
experimental data that is marked with the warning comment `W'.}
This shows the validity of approximation \eqref{eq:fitsigmainel}.

\subsection{Secondary multiplicity}
Bubble chamber and accelerator experiments  
have shown that the number of charged particles 
created in proton -- (anti)proton collisions, i.e. the charged multiplicity, increases as a function of the
incident proton energy.\footnote{The charged particle multiplicity
in $pp$ and $p \bar{p}$ interactions is virtually identical at ISR energies
$\sqrt{s} = 53 $ GeV (Ref. \cite{Breakstone:1983pb}; see also 
Ref. \cite{CollinsMartin}).}  
We find that up to the highest energies currently accessible, $\sqrt{s} \leq \EE{1.8}{3} $ GeV,  the charged particle
multiplicity in non-diffractive $pp$ interactions is well fitted with
\beqs
\label{eq:theory:chmult}
\mathcal{M}^{\textrm{fit}}_{\textrm{ch}} (s) = 0.89 + 1.24 \log s + 0.34 \log^2 s  + 0.077 \log^3 s \, .
\eeqs
This functional form is an extention of an approximation
due to Matthiae \cite{Matthiae:1983ke} (see also Ref. \cite{CollinsMartin}) which is valid only up to
$\sqrt{s} \leq 540 $ GeV.
The last logarithmic term, which does not appear in the approximation by Matthiae,
is required in order to fit the multiplicity at both high and low energies.
We present in figure \ref{fig:mult} experimental data\footnote{The experimental data is taken from
Refs. \cite{Ansorge:1988kn, Alpgard:1982kx,Benecke:1974if, Biyajima:2001ud, Breakstone:1983ns,Morse:1976tu}.
The data at $\sqrt{s}=\EE{1.8}{3}$ GeV is obtained by the 
E735 experiment \cite{Lindsey:1991pt}, which does not cover the full particle
phase space. We use results from Ref. \cite{Biyajima:2001ud}, who have determined
the charged particle multiplicity through a fit to experimental data on the
multiplicity distribution.} 
together with the approximation given in eq. \eqref{eq:theory:chmult} and
\texttt{PYTHIA} results (see section \ref{sec:results} below).

A logarithmic dependence $\mathcal{M}_{\textrm{ch}} \propto \log s$
is commonly interpreted as due
to an increase in phase space because 
the range of allowed rapidities scales as $\log (s/m^2 c^4)$ (e.g., Ref. \cite{CollinsMartin}).
A stronger increase in secondary multiplicity
is then attributed to an additional rise in the level
of the observed central rapidity plateau, the origin of which 
is not understood from first principles.

At high energies, data on the neutral particle multiplicity is sparse because
of experimental difficulties. As a result, there is no fit to the neutral particle multiplicity
that extends to $\sqrt{s}  \gtrsim 50$ GeV which is based on experimental data.
A fit to the separate multiplicities
of both charged and neutral pions and charged kaons created in $pp$ collisions for center-of-mass
energies \mbox{$\sqrt{s} < 53$ GeV} was presented in Ref. \cite{Antinucci:1972ib}.

The scarcity of experimental data on separate particle multiplicities at high
energies motivates the use of event generators such as \texttt{PYTHIA}.
In section \ref{sect:numcomp}, we show  that \texttt{PYTHIA} correctly reproduces
experimental results on the total charged multiplicity.
In section \ref{sec:res:secenmult},
we present a fit to \texttt{PYTHIA} results on charged and neutral
pion and kaon multiplicities in the energy range 
\mbox{$ 43 \textrm{ GeV} < \sqrt{s} < \EE{1.4}{3} \textrm{ GeV} $}.

\section{Kinematics and secondary particle distribution}
\label{sec:cross_kin}
In this and the following sections, we consider an energetic proton that moves along
the $z$-axis and collides with a 
proton at rest, i.e. a fixed target. This scattering geometry is referred to as the lab frame.
We use $p_z$ to denote a longitudinal momentum, along the $z$-axis, 
and $p_T$ to denote
a transverse momentum.

\subsection{Kinematics}
Assuming that the secondary particle distribution is symmetric around the collision
axis, the phase space of the outgoing particles is fully parameterized with two independent 
kinematical variables.
Here, we choose the energy $\ep$ and the rapidity $y$, which is defined as
\beqs
\label{kin:defrap}
y = \frac{1}{2} \ln \lb \frac{\ep + p_z c}{\ep - p_z c} \rb 
\quad \Leftrightarrow \quad \tanh
y = \frac{p_z c}{\ep} \, .
\eeqs
For given particle energy $\ep$, the rapidity cannot take any value.
The mass-shell relation implies that \mbox{$-y_1 < y < y_1$}, where
\beqs
\label{eq:kin:yminmax1}
y_1 = \textrm{arccosh} \lb \frac{\ep}{m c^2} \rb  \, ,
\eeqs
and $m$ is the secondary (pion or kaon) mass. A second requirement
follows from energy conservation in the $pp$ collision.
If the energy of the secondary particle $\ep > m_p c^2$, the rapidity 
is additionally bounded by $y > y_2$, where
\beqs
\label{eq:kin:yminmax2}
y_2 = \textrm{arctanh} \lb \frac{1}{\beta'_p} -\frac{2 m_p c^2}{ \beta'_p \ep} \rb  \, .
\eeqs
In this equation $\beta'_p$ is the proton velocity in the center-of-mass frame
in units of $c$, which we take to be equal to one for incident proton energies
$E_p \gg m_p c^2$ in the following calculations.
Note that eq. \eqref{eq:kin:yminmax1} can be applied in any frame, while
eq.  \eqref{eq:kin:yminmax2} only holds in the lab frame.

\subsection{Secondary particle distribution}
\label{sec:dist}
We are interested in the particle distribution for one-pion and one-kaon inclusive
processes,
\beqs
p p \to X Y \, ,
\eeqs
where $X$ denotes a single pion or a single kaon and $Y$ may be any combination
of particles with the appropriate quantum numbers.
We denote by $n(\ep,y) d\ep dy$ the number of created  particles of a given species with energy and rapidity
in the range \mbox{($\ep \ldots \ep+ d\ep$)} $\times$ \mbox{($y \ldots y + d y$)}:
\beqs
n (\ep,y) =  \frac{d^2 N}{d \ep dy} = \frac{1}{\sigma_{\textrm{inel}}}
\frac{d^2 \sigma}{d \ep dy} \, ,
\eeqs
where $\sigma_{\textrm{inel}} = \sigma_{\textrm{nd}}  + \sigma_{\textrm{sd}}  + \sigma_{\textrm{dd}}$ 
is the inelastic $pp$ cross section and $\sigma$ is the inclusive cross section to 
detect a particle of a given kind (assuming an ideal detector). This cross section is equal to the
weighted sum of $n$-particle exclusive cross sections 
$\sigma_n$ (i.e., the cross section to create exactly $n$ particles)\footnote{\label{footnote:KNO}We do
not consider the exclusive cross sections separately because we are interested in 
particle creation by all processes together.
To a first approximation, the relative sizes of 
the $n$-particle exclusive cross sections depend on energy only
through the total multiplicity \cite{Koba:1972ng}. This `KNO scaling'
is known to be violated at energies $\gtrsim$ 500 GeV \cite{Alpgard:1982kx,Alner:1984is}.}:
\beqs
\sigma = \sum_n n \sigma_n = \mathcal{M} \sigma_{\textrm{inel}} \, ,
\eeqs
where $\mathcal{M} = \mathcal{M} (s) $ is the  multiplicity of the given particle species.
The particle distribution $n(\ep,y)$ is related to the Lorentz invariant differential
cross section $\ep \, d^3 \sigma  / dp^3$, which is often used to represent experimental
data, as follows:
\beqs
\label{eq:theory:n}
n(\ep,y)  = \frac{2 \pi}{c}  \lb  \frac{ m^2 + p_T^2}{\sigma_{\textrm{inel}} \ep}\rb  \lb  \ep \frac{d^3 \sigma}{d^2 p_T d p_z} \rb \, . 
\eeqs

\section{Numerical method}
\label{sec:method}

\subsection{Configuration of the \texttt{PYTHIA} event generator and initial conditions }
\label{section:configuration}
The $pp$ interaction is simulated
with \texttt{PYTHIA} version 6.324 using default values for most of the control parameters.
Elastic and diffractive processes are included by 
selecting \texttt{MSEL=2}. In comparing \texttt{PYTHIA} results to experimental data
on the cross section and charged multiplicity, we allow for pion and kaon decay.
In determining the parameterizations of the particle distributions,
pion and kaon decay are switched off with
the command \texttt{MDCY(PYCOMP(ID),1)=0}, where \texttt{ID} is 
the corresponding particle identification number.
This approach separates the physics in the $pp$ collision from subsequent processes,
such as secondary synchrotron emission prior to decay, etc.

The \texttt{PYTHIA} code relies on
the Lund string model \cite{Andersson:1983ia} to determine the
distribution and multiplicity of secondary particles.
We present a brief discussion of this model in appendix
\ref{sec:applund};
more details can be found in the \texttt{PYTHIA}
manual \cite{Sjostrand:2003wg}.

We simulate $pp$ collisions for incident proton energies
$E_p = 10^3$ GeV, $E_p = 10^4$ GeV, $E_p = 10^5$ GeV and $E_p = 10^6$ GeV colliding with a proton
at rest. For higher values of the incident proton energy, \texttt{PYTHIA}
signals a loss of accuracy in kinematical variables in some of the generated
events.

\subsection{Fitting procedure}

The secondary particle distributions are discretized, spanning the full range of available
energy and kinematically allowed rapidity. In this process, the energy
is divided into 
200 bins with size $\Delta \ep_i$ with a logarithmic division and
the rapidity is divided into 100 bins with size $\Delta y_i$ with a linear division.
The logarithmic
energy division is chosen because we consider up to seven
 energy decades; the rapidity division is linear because the range of allowed
rapidities scales with the logarithm of energy.
The number of bins is limited by computational issues, as
data files become increasingly large and fitting becomes increasingly
time-consuming with an increasing number of bins.
We have verified that this number of bins is sufficient for convergence of
the resulting parameterization.

We use \texttt{MINUIT}\footnote{CERN Program Library entry D506;
documentation is available on the website \texttt{http://wwwasdoc.web.cern.ch/wwwasdoc/minuit/}} as a minimization
algorithm for the weighted squared difference
between the \texttt{PYTHIA} results and the particle distribution fit function
$n(\ep,y)$.\footnote{We do not explicitly write the dependence of $n$ on $E_p$ here and in 
the following sections.}
We consider only statistical errors in the \texttt{PYTHIA} results.
We simulate $N_{\textrm{ev}} = 10^6$ collisions for every incident proton energy,
which results in a statistical error of a few percent near the maximum values
of $\ep \, n(\ep,y)$.
We compare our results with parameterizations based on other event generators 
to obtain an estimate of the importance of systematic uncertainties within the models
underlying \texttt{PYTHIA}  in section \ref{sec:disc:val}.

The relative deviation between a \texttt{PYTHIA} data point $n_i$ and the fitted value $n(\ep_i,y_i)$
is expressed as
\beqs
\delta_i = \frac{n(\ep_i,y_i) - n_i}{n_i}  \, ,
\eeqs
where $n_i$ is the number of particles in a bin with average energy
$\ep_i$ and average rapidity $y_i$ divided by the bin size $\Delta \ep_i \times \Delta y_i$.
We note that the deviations are expected to follow a Gaussian distribution
with average value
\beqs
\label{fit:err}
\langle \delta_i \rangle &\propto & \sqrt{\frac{1}{\ep_i n_i}} \, ,
\eeqs
where the dependence on energy is due to the logarithmic energy binning. 
In particular,  the average deviation size is expected to be roughly independent of energy 
for a $n(\ep) \propto \ep^{-1}$  energy spectrum.

\section{Results}
\label{sec:results}

\subsection{Comparison of \texttt{PYTHIA} results with experimental data}
\label{sect:numcomp}
We show in figures \ref{fig:sigma} and  \ref{fig:mult} 
experimental data on the $pp$ cross section
and charged multiplicity together with
\texttt{PYTHIA} results.
In producing these figures, 
we have not switched off any natural particle
decays in the \texttt{PYTHIA} simulations in order to compare
\texttt{PYTHIA} results with experimental data (cf. section \ref{section:configuration}).

We observe from figure \ref{fig:sigma} that the  \texttt{PYTHIA} cross
sections are compatible with an extrapolation of the experimental data
(see footnote \ref{footnote:PPDS}) in the energy range \mbox{$  43 \textrm{ GeV} < \sqrt{s} < \EE{1.4}{3} \textrm{ GeV} $}.
In figure \ref{fig:mult}, we show a comparison between experimental data and 
\texttt{PYTHIA} results on the charged multiplicity. In this work, we are interested in particle
creation by all inelastic processes. However, experimental data 
on the charged multiplicity resulting from all inelastic processes is available only up
to $\sqrt{s} = 62$ GeV \cite{Benecke:1974if, Breakstone:1983ns, Morse:1976tu}. Experimental data on
the charged multiplicity resulting from the restricted class of non-single-diffractive (NSD) interactions
is available up to much higher energies  $\sqrt{s} = \EE{1.8}{3}$ GeV 
\cite{Alner:1985wj,Ansorge:1988kn,Biyajima:2001ud,Breakstone:1983ns}.
To verify our numerical method, we have performed a separate
simulation\footnote{For the NSD case, we have switched off single diffraction in the
event generation with the commands \texttt{MSUB(92)=0} and \texttt{MSUB(93)=0}.}
of  NSD interactions to compare the NSD charged multiplicity with experimental data. We 
show in figure \ref{fig:mult} that the \texttt{PYTHIA} results on the charged
multiplicity due to both inelastic
processes and NSD processes  are compatible with experimental data.

\subsection{Average secondary energy and multiplicity}
\label{sec:res:secenmult}
\begin{table}
\caption{\label{tab:energetics} Average fraction of the incident proton energy carried by the outgoing particle species.}
\begin{ruledtabular}
\begin{tabular}[c]{c c c c c c c c c c c}
& $N_p $ & $N_s$  & $\gamma$ &  ${\pi^+}$ & ${\pi^-}$ & ${\pi^0}$ & ${K^+}$ & ${K^-}$ & ${K^0}$ & ${\overline{K^0}}$ \\
\hline
$f$ &  0.56  & 0.033  & 0.013  & 0.13 & 0.095 & 0.12 & 0.016 & 0.011 & 0.014 & 0.011  \\
\end{tabular}
\end{ruledtabular}
\end{table}

\begin{table}
\caption{\label{tab:multiplicity} Numerical values for the constants in the multiplicity approximation formula \eqref{eq:res:Mapp}.}
\begin{ruledtabular}
\begin{tabular}[c]{c c c c c c c c}
& $ \pi^+  $ &  $\pi^- $ & $\pi^0 $ & $ K^+ $ & $ K^-$ & $K^0$ & $\overline{K^0}$ \\
\hline
$c_0$  & $  4.5  $ & $  3.8 $ & $ 4.9 $ &    $ 0.49  $ & $ 0.32 $     & $ 0.36   $ & $ 0.29 $ \\
$c_1$  & $  -1.7  $ & $  -1.7 $ & $ -2.1 $ & $ -0.23  $ & $ -0.20 $ & $ -0.17 $ & $ -0.18 $  \\
$c_2$  & $ 0.50  $ & $ 0.50  $ & $ 0.58  $ &    $ 0.063  $ & $ 0.060 $   & $ 0.054  $ & $ 0.054 $ \\
\end{tabular}
\end{ruledtabular}
\end{table}
The fraction $f$ of the incident proton energy 
carried by a certain secondary particle species 
is virtually independent of the incident proton energy (see Ref. \cite{Feynman:1969ej}).
The average fractions for nucleons, photons, pions and kaons found by \texttt{PYTHIA} simulations 
are given in table \ref{tab:energetics}.
In this table, $N_p$ and $N_s$ denote primary and secondary nucleons, respectively (see below).
Other possible secondaries (direct electrons, muons, neutrinos) together
carry less than $0.1$$\%$ of the incident proton energy.

We define the primary nucleon as the most energetic 
outgoing nucleon. The probability that the primary nucleon
is a proton is $0.70$; if it is a proton, it carries
an average fraction $0.63$ of the incident proton energy. The probability that the primary nucleon
is a neutron is $0.30$; if this is the case, the average energy fraction is $0.41$. The
energy fraction carried by the
primary nucleon as shown in table \ref{tab:energetics} represents the weighted average.

We fit \texttt{PYTHIA} results on the secondary particle multiplicities 
within the energy range \mbox{$  43 \textrm{ GeV} < \sqrt{s} < \EE{1.4}{3} \textrm{ GeV} $}.
We find that both charged and neutral pion and kaon multiplicities are well approximated
with the following function:
\beqs
\label{eq:res:Mapp}
\mathcal{M}_{i} = c_0 + c_1 \log s + c_2 \log^2 s \, ,
\eeqs
where $c_0$, $c_1$ and $c_2$ are numerical constants whose values are given in 
table \ref{tab:multiplicity} and $s$ is expressed in units of GeV$^2$.
The charged kaon multiplicity deduced from eq. \eqref{eq:res:Mapp} is within $\sim$$5\%$ 
of experimental data at $ \sqrt{s} = 45 \textrm{ GeV} $ \cite{Antinucci:1972ib}.
The charged pion multiplicities determined by this equation are $\sim$$10\%$ lower than the experimental values.
This discrepancy can be attributed to the fact that we have considered only prompt pions (i.e., excluding
pions from kaon decay).

\subsection{Pion and kaon energy spectra}
\label{sect:res:spec}
\begin{figure}
\begin{center}
\includegraphics[angle=270, width=8.6cm]{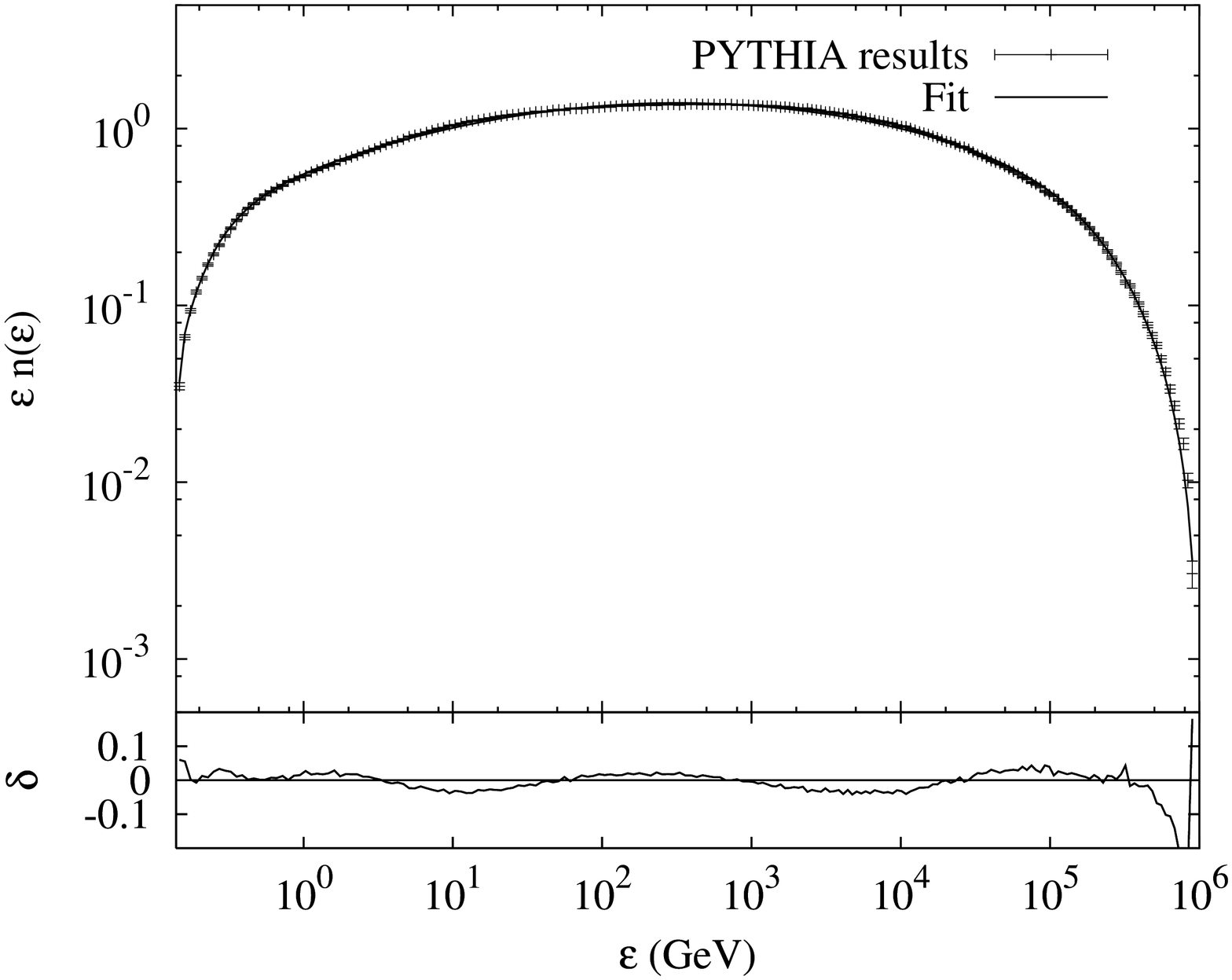}
\hspace{0.5cm}

\includegraphics[angle=270, width=8.6cm]{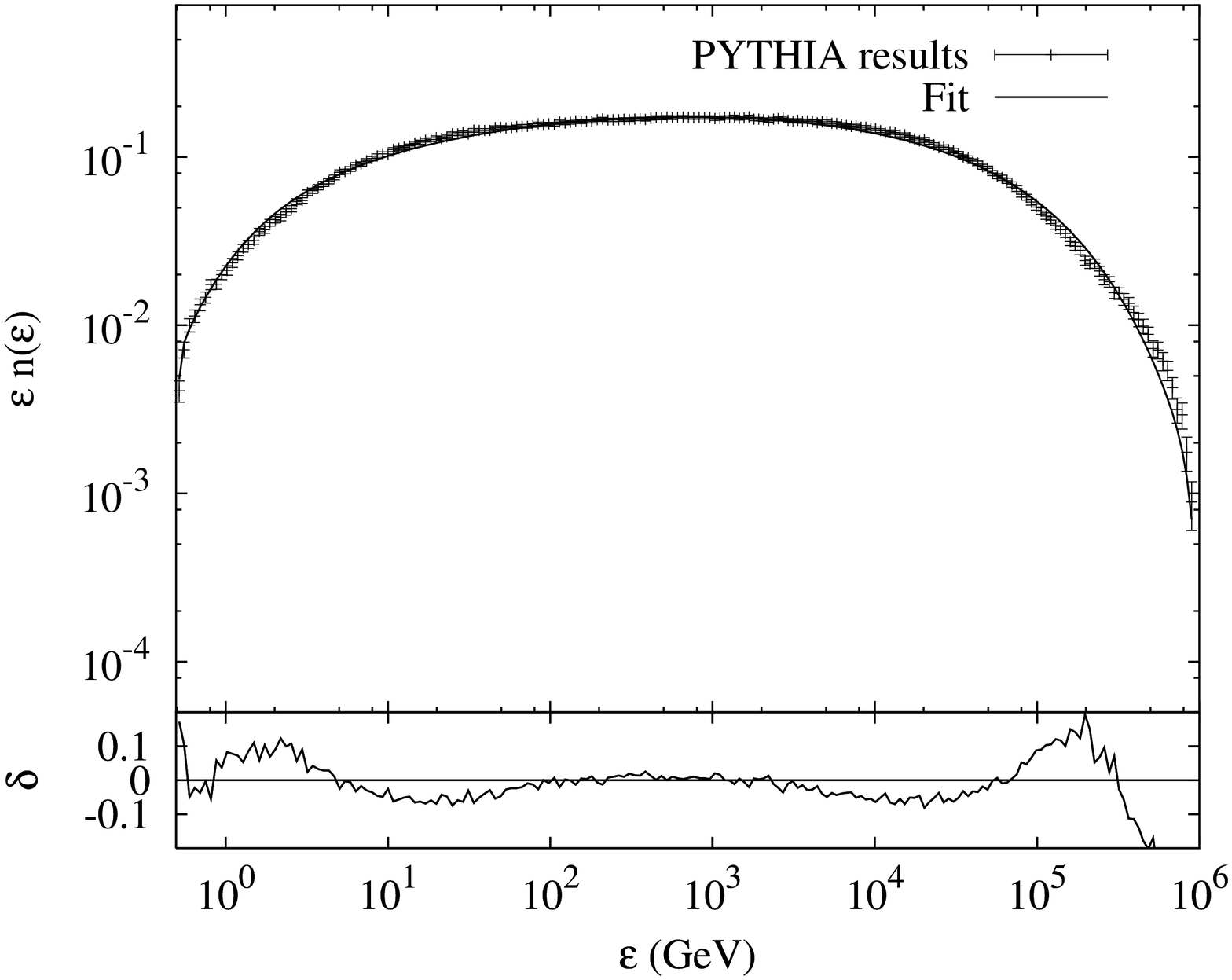}
\end{center}
\caption{Energy spectra of $\pi^+$ (top panel) and $K^+$ (bottom panel) mesons created in 
a collision with incident proton energy \mbox{$E_p = 10^6$} GeV.
Note the different scales on the vertical axes.
Upper graphs: comparison of \texttt{PYTHIA} results and fit to the energy spectrum normalized
as $\ep \, n(\ep)$;
lower graphs: deviation \mbox{$\delta_i = n(\ep_i,y_i)/n_i -1$} between \texttt{PYTHIA} results points
and fitted values.}
\label{fig:res:spec}
\end{figure}
We present in figure \ref{fig:res:spec} the $\pi^+$ and $K^+$ energy spectra resulting from a collision
of a proton with incident energy $E_p = 10^6$ GeV with a proton at rest. 
We find that the energy spectra for all secondary
particles and incident proton energies \mbox{$10^3$ GeV $ < E_p < 10^6$ GeV} are similar in
shape.\footnote{Figures of these and others fits are available at \texttt{http://www.nikhef.nl/\~{ }hkoers/ppfit}.}
 To a first approximation, the 
energy spectra follow a 
$\ep^{-1}$ power-law, reflecting the absence of an energy scale between the secondary mass
and the maximum available energy. This is supplemented with
additional functions that we denote with $\alpha (\ep)$, $\beta(\ep)$, $\gamma_1(\ep)$
and $\gamma_2(\ep)$
(here and in the following we do not explicitly write the dependence of the
model parameters on the incident proton energy $E_p$ to avoid cluttering of the
notation). Thus we write the pion and kaon energy spectrum in the following way:
\beqs
\label{eq:spec:res}
n(\ep) = n_0   \ep^{-1} \alpha (\ep) \beta(\ep) \gamma_1 (\ep)  \gamma_2 (\ep)\, ,
\eeqs
where $n_0$ is a normalization constant,
$\alpha (\ep)$ accounts for the convex shape on a log-log scale,
$\beta (\ep)$ incorporates an exponential
decline at higher and lower energies, $\gamma_1 (\ep)$ is a strong
cutoff near the mass threshold and  $\gamma_2 (\ep)$ is a strong
cutoff near the maximum available energy.
These functions are parameterized as follows:
\bsub
\label{eq:res:spec1}
\beqs
n_0& = &  1.21 \times 10^{p_0 + p_1 p_2^2} \, ; \\
\alpha (\ep) & = &  \ep^{p_1 (\log (\ep) - 2 p_2 )}  \, ; \\
\beta (\ep) & = &  10^{- (\ep/\ep_3)^{p_{30}}} 10^{- (\ep/\ep_5)^{p_{50}}}   \, ; \\
\gamma_1 (\ep) & = & \tanh \lb p_{70} \log \lb \ep / m c^2 \rb \rb   \, ; \\
\gamma_2 (\ep) & = & \tanh \lb p_{80} \log \lb E_p / \ep \rb \rb  \, ,
\eeqs
\esub
where $\ep_3 \equiv 10^{p_2 + p_4}$, $\ep_5 \equiv 10^{p_2 + p_6}$ and all energies
are expressed in units of GeV. The following parameters vary with incident proton energy:
\bsub
\label{eq:res:spec2}
\beqs
p_0 & = & p_{00} + p_{01} \log (E_p) \, ; \\
p_1 & = & p_{10} + p_{11} \log (E_p) \, ; \\
p_2 & =&  p_{20} + p_{21} \log (E_p) \, ;  \\
p_4 & = & p_{40} - p_2 \, ; \\
p_6 & = & p_{60} + p_2 \, .
\eeqs
\esub
Thus, the energy spectrum of secondary pions and kaons is fully described 
in terms of 12 free parameters $p_{ij}$ for every particle species. 
These parameters and their numerical values, which are determined by a least-squares fit,
are given in  table \ref{table:num:spec}.

For pions, deviations between fit values $n(\ep_i)$ and \texttt{PYTHIA} results 
$n_i$  are less than 5\% except for very high energies ($\ep \geq E_p /2$)
and some occasional points near the mass threshold where the deviation is
$\sim$10\%. For kaons, statistical fluctuations are larger since the number of
kaons to pions is roughly 1:10. At intermediate energies the fit is 
nevertheless within $\sim$5\% of \texttt{PYTHIA} results except for some isolated points.
Near the mass threshold deviations increase
to $\sim$20\%; at very high energies, where $\ep \, n(\ep,y)$ is typically more than an order
of magnitude smaller than its maximum value, deviations can increase to $\sim$40\%.
We have verified that the parameterized spectra integrate to the right multiplicities as given in eq. 
\eqref{eq:res:Mapp} within a few percent, except for the $K^0$ spectrum 
for which the deviation is $\sim$$10$\% at the low end
of the simulated proton energy range.

\subsection{Pion and kaon energy and rapidity distributions}
We present the pion and kaon 
rapidity distributions, i.e. $n(\ep,y)$ at fixed $\ep=\ep_0$, for incident proton
energy $E_p = 10^6$ GeV and secondary particle energy \mbox{$\ep = 10^3$ GeV}
in figure \ref{fig:fit:dist}.
We find that rapidity distributions 
for different proton energies and different secondary particle energies are very similar in shape.
This shape is different for pions and for kaons, hence
we treat pions and kaons separately in the following.

\subsubsection{Pions}
\label{sect:res:dist}
\begin{figure}
\begin{center}
\includegraphics[angle=270, width=8.6cm]{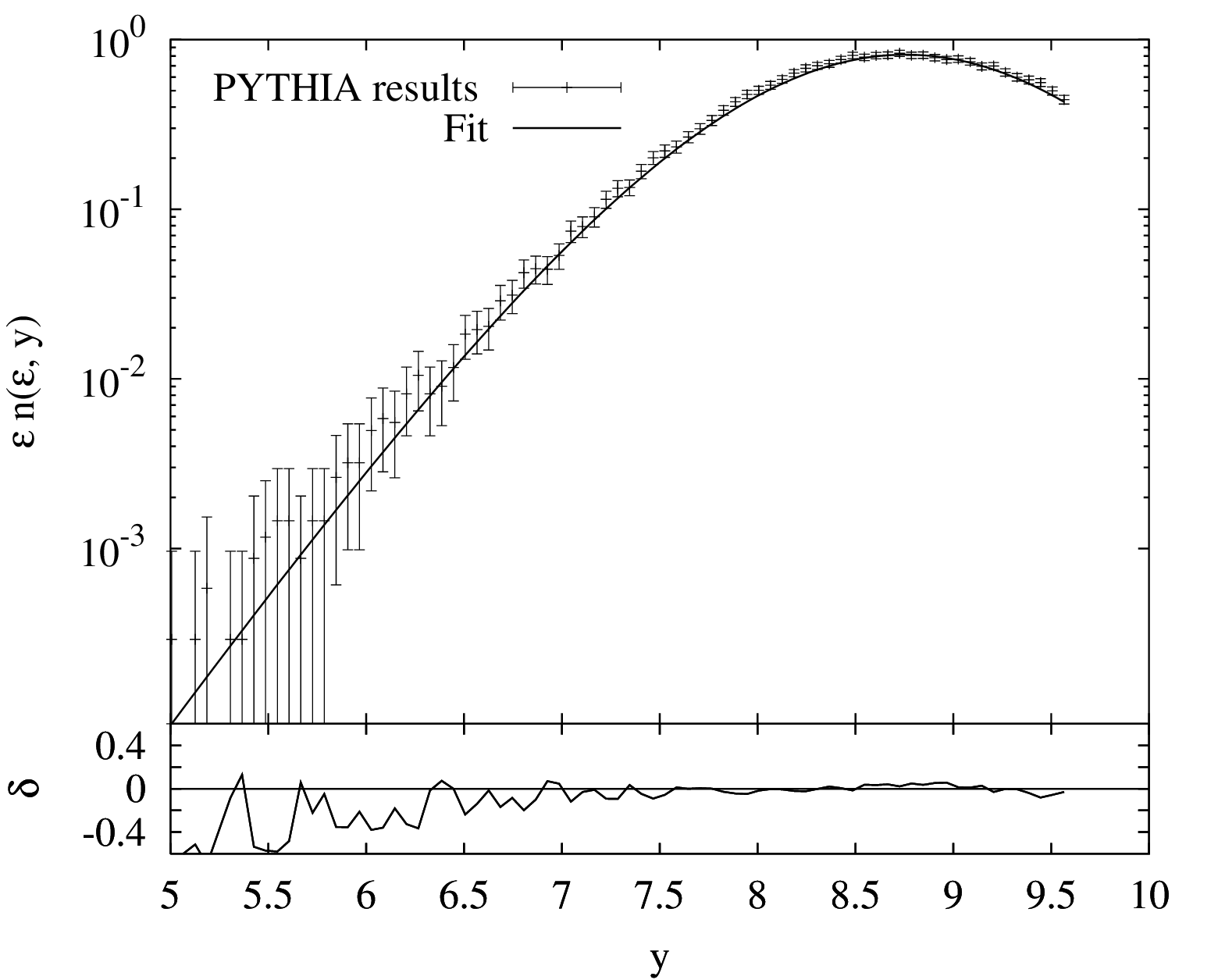}
\includegraphics[angle=270, width=8.6cm]{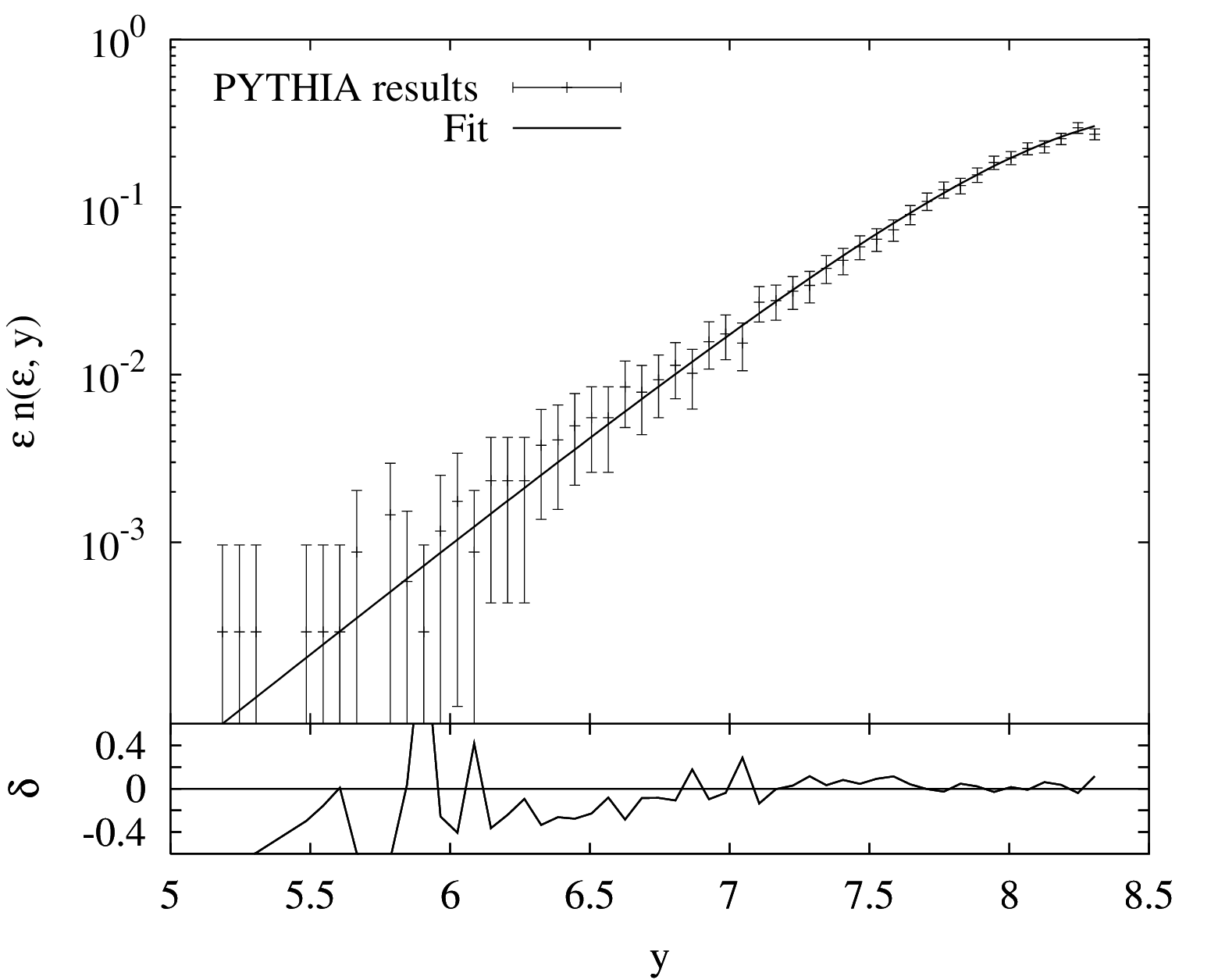}
\end{center}
\caption{
Rapidity distributions of $\pi^+$ (top panel) and $K^+$ (bottom panel) secondaries created in 
a collision of a proton with energy \mbox{$E_p = 10^6$} GeV with a proton at rest.
The secondary particle energy is taken to be $\ep = 10^3$ GeV.
Upper graphs: comparison of \texttt{PYTHIA} results and fit to the rapidity distribution
$\ep \, n(\ep,y)$ as a function of $y$;
lower graphs: deviation \mbox{$\delta_i = n(\ep_i,y_i)/n_i -1$} between  \texttt{PYTHIA} results
and  fitted values.}
\label{fig:fit:dist}
\end{figure}
The pion rapidity distributions at fixed energy are found to be approximately Gaussian near their
maximum values (see fig. \ref{fig:fit:dist}). At intermediate pion energies, $\ep \sim \sqrt{E_p}$,
the distributions exhibit a low-rapidity tail that falls off exponentially
(all energies are expressed in units of GeV). 
The distributions fall off very steeply at the boundaries of the kinematical domain
given in eqs. \eqref{eq:kin:yminmax1} and  \eqref{eq:kin:yminmax2}.

We factorize the full particle distribution $n(\ep,y)$ 
into a modified energy spectrum $\tilde{n} (\ep)$ 
and a rapidity-dependent function $ \phi(\ep,y)$ that contains both a Gaussian and an exponential part:
\beqs
\label{eq:res:dist:pion}
n_\pi (\ep,y)  =  \tilde{n}(\ep) \phi (\ep,y) \,  ,
\eeqs 
where
\bsub
\label{eq:res:dist2}
\beqs
\tilde{n} (\ep) & = & n_0  \ep^{-1} \alpha (\ep) \beta(\ep) \gamma_2 (\ep)  10^{q_0 + 2 q_3} \, ; \\
\phi(\ep,y) & = & 10^{ - 2 \sqrt{q_3 q_1^2 (y-q_2)^2 + q_3^2}} \, ,
\eeqs
\esub
and $n_0$, $\alpha (\ep)$, $\beta (\ep)$ and $\gamma_2(\ep)$ are defined in eqs. \eqref{eq:res:spec1}.
The parameters $q_i$ depend on the pion
energy $\ep$ and on the incident proton energy $E_p$ in the following way
(here and in the following we do not explicitly write the dependence of the parameters
$q_i$ on the pion and proton energies to avoid cluttering of notation):
\bsub
\label{eq:res:qpar}
\beqs
q_0 & = & q_{00} + q_{01} \xi^2 \, ; \\
q_1 & = & q_{10} + q_{11} \lb \xi + q_{12} \rb ^2 + q_{13} \xi^4 \, ; \\
\nn q_2 & = & \ln (\ep) +  q_{20} +  q_{21} \lb \log(\ep)+1 \rb - 10^{q_{22} \log (E_p) \log \lb \ep / E_p \rb} \\
&  & \, \, + \, 10^{q_{23} \log (E_p) \log \lb m / \ep \rb}   \, ;  \\
q_3 & = & q_{30} + 10^{q_{31} + q_{32} \log (\ep) } \, ,
\eeqs
\esub
where we introduced the variable
\beqs
\xi \equiv  2 \log (\ep) / \log (E_p) -1 \, .
\eeqs
Hence, we have parameterized the pion energy and rapidity distributions in terms of 24 free parameters
$q_{ij}$ for every pion species ($\pi^+$, $\pi^-$ and $\pi^0$). The fitted values for the coefficients are given in table \ref{table:num:dist}.

Deviations between the parameterizations and \texttt{PYTHIA} results are within 10\%
in the range in which the rapidity distribution is within one order of magnitude of the maximum value
and for pion energies  \mbox{1 GeV $< \ep <  0.1 \, E_p$},
except for a few isolated points that are typically within 20\%.
At high energies, \mbox{$\ep \geq 0.1 \, E_p$}, deviations increase
to $\sim$30\% at the borders of the considered rapidity interval,
in concordance with eq. \eqref{fit:err}.

We have verified that integrating the energy and rapidity distributions over rapidity reproduces
the energy spectra. The deviations between these spectra and \texttt{PYTHIA} results
are similar to the deviations for the direct fit to the energy spectra
(see section \ref{sect:res:spec}), except at very low energies $\ep \lesssim 2 m_\pi$ where
deviations increase to $\sim$30\%. The multiplicities obtained by integrating the
distributions over energy and rapidity are within a few percent of those given by
eq. \eqref{eq:res:Mapp}.

\subsubsection{Kaons}
The shape of the kaon rapidity distributions is similar to the low-rapidity part of the
pion rapidity distributions (see fig. \ref{fig:fit:dist}). We find
that the kaon energy and rapidity distributions are well described with:
\beqs
\label{eq:res:dist:kaon} 
n_K(\ep,y)  =   \bar{n}(\ep) \phi (\ep,y) \, ,
\eeqs 
where $\phi (\ep,y)$ is defined in terms of model parameters $q_i$ in
eqs. \eqref{eq:res:dist2} and $ \bar{n}(\ep)$ is a modified energy spectrum:
\beqs
\bar{n} (\ep) =  n_0  \ep^{-1} \alpha (\ep) \beta(\ep) 10^{q_0 + 2 q_3} \, . 
\eeqs
The quantities $n_0$, $\alpha (\ep)$ and $\beta (\ep)$ are defined in eqs. \eqref{eq:res:spec1}.
We find that the parameterizations for $q_i$ given in eqs. \eqref{eq:res:qpar} approximate the \texttt{PYTHIA} results
well if we  fix $q_{23} =0$. Therefore, the kaon energy and rapidity distributions are fully parameterized in terms of
23 free parameters for every kaon species $(K^+, K^-, K^0$ and $\overline{K^0})$. The fitted values for these parameters are  presented in table \ref{table:num:dist}.

Deviations between the approximation \eqref{eq:res:dist:kaon} and \texttt{PYTHIA} results are
similar to the deviations for the parameterizations of the pion distributions,
except that fluctuations are larger. This results in deviations up to $\sim$30\% at isolated points for all energies.

For the $K^-$ and $\overline{K^0}$ mesons,
integrating the full distributions over the rapidity reproduces the energy spectra with
deviations similar to those for the direct parameterizations of the energy spectra
presented in eq. \eqref{eq:spec:res}. For $K^+$ and $K^0$ mesons, deviations
near the mass threshold are $\sim$30\%, while at very high energies ($\ep \gtrsim E_p/2$)
the deviations can increase to $\sim$50\%. These large deviations occur only
at energies where $\ep \, n(\ep,y)$ is more than an order
of magnitude smaller than the maximum value. The multiplicities obtained by integrating the
parameterized distributions \eqref{eq:res:dist:kaon} 
over energy and rapidity are within a few percent of the values given by
eq. \eqref{eq:res:Mapp}.

\section{Applications}
\label{sec:app}

In this section, we consider applications of the parameterized particle distributions presented in eqs.  \eqref{eq:res:dist:pion} 
and   \eqref{eq:res:dist:kaon}.
We show explicitly how the parameterizations
can be used to study
correlations between the energy and outgoing
angle of secondary particles
for a general scattering geometry, viz. two
protons with different energies colliding at an arbitrary angle.
We also present an example in which the
energy spectrum of photons due to decay of  $\pi^0$ 
mesons created in a $pp$ interaction is derived from the parameterized  $\pi^0$
distribution.
For clarity, we consider only $\pi^0$ mesons in this section,
but the presented methods  are applicable to all pions
and kaons.

\subsection{Head-on proton -- proton collision}

\subsubsection{Full secondary particle distribution}
\begin{figure}
\begin{center}
\includegraphics[angle=270, width=8.6cm]{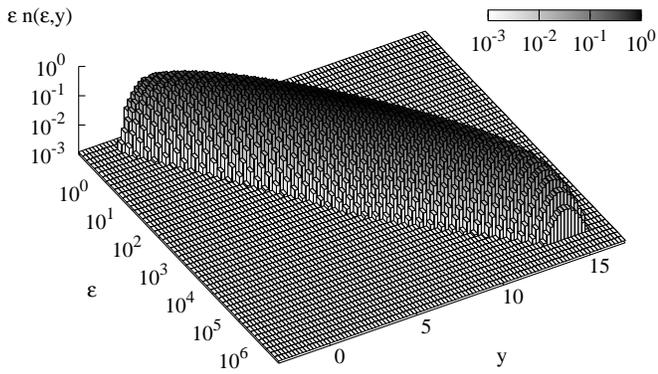}
\end{center}
\caption{The $\pi^0$ distribution $\ep \, n(\ep,y)$ as a function of energy $\ep$
and rapidity $y$ after a collision of a proton with energy  $10^6$ GeV with
a proton at rest (lab frame). The discretization and the observed `floor'
are for presentational purposes.}
\label{fig:app:lab}
\end{figure}
\begin{figure}
\begin{center}
\includegraphics[angle=270, width=8.6cm]{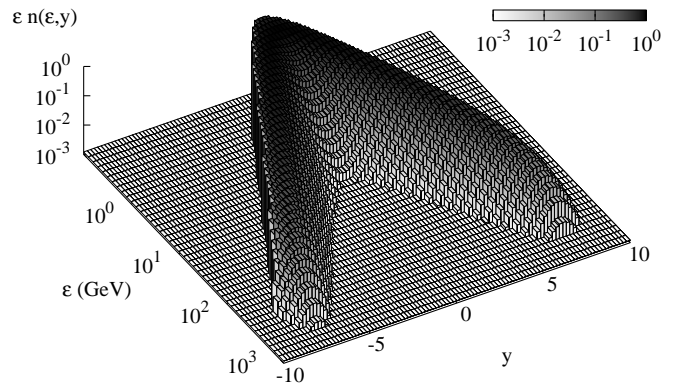}
\end{center}
\caption{
The $\pi^0$ distribution $\ep \, n(\ep,y)$ as a function of energy $\ep$
and rapidity $y$ after a collision of two protons  with energy  $730$ GeV (center-of-mass frame).
The discretization and the observed `floor' are for
for presentational purposes.}
\label{fig:app:com}
\end{figure}
\begin{figure}
\begin{center}
\includegraphics[angle=270, width=8.6cm]{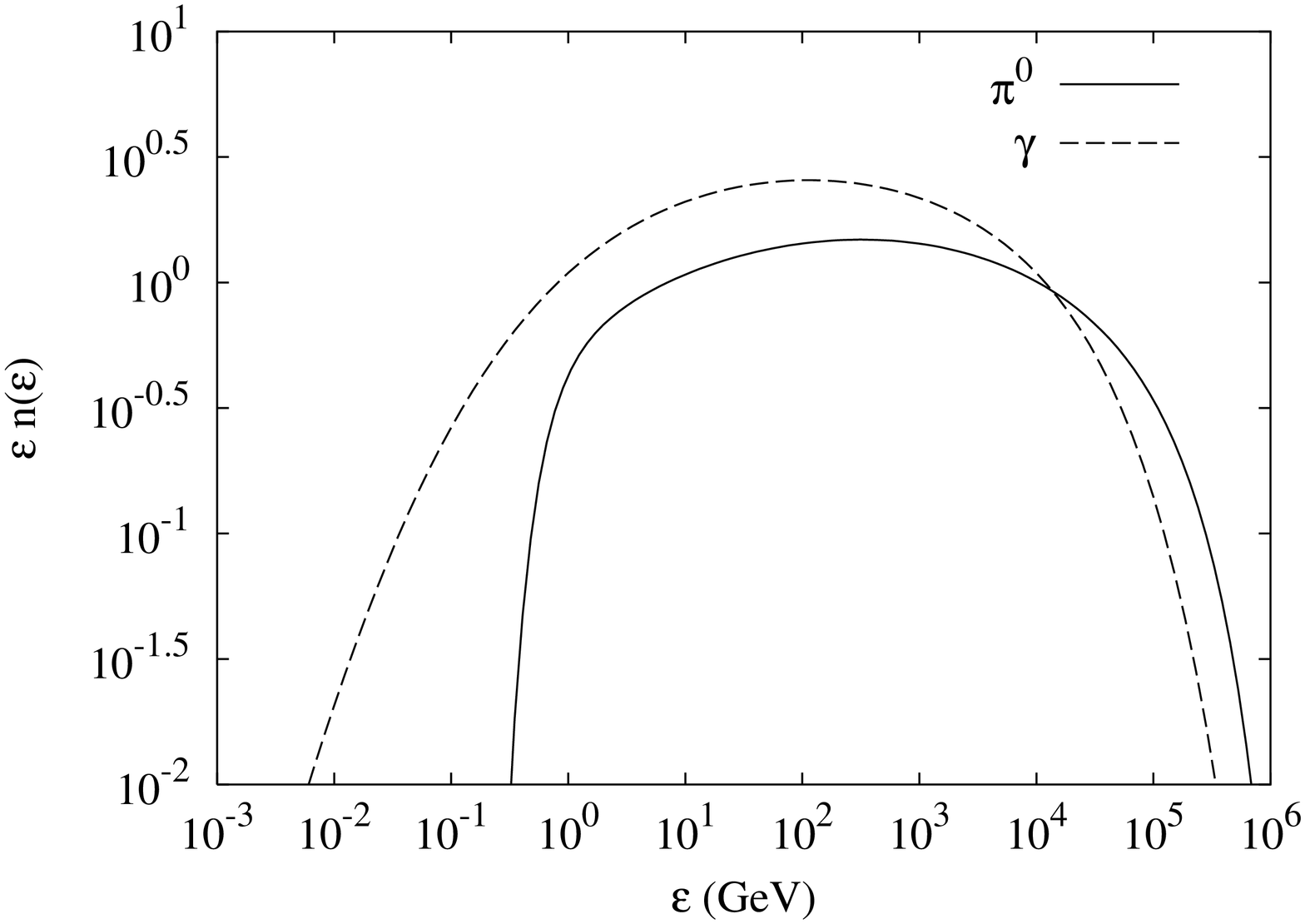}
\hspace{0.5cm}

\includegraphics[angle=270, width=8.6cm]{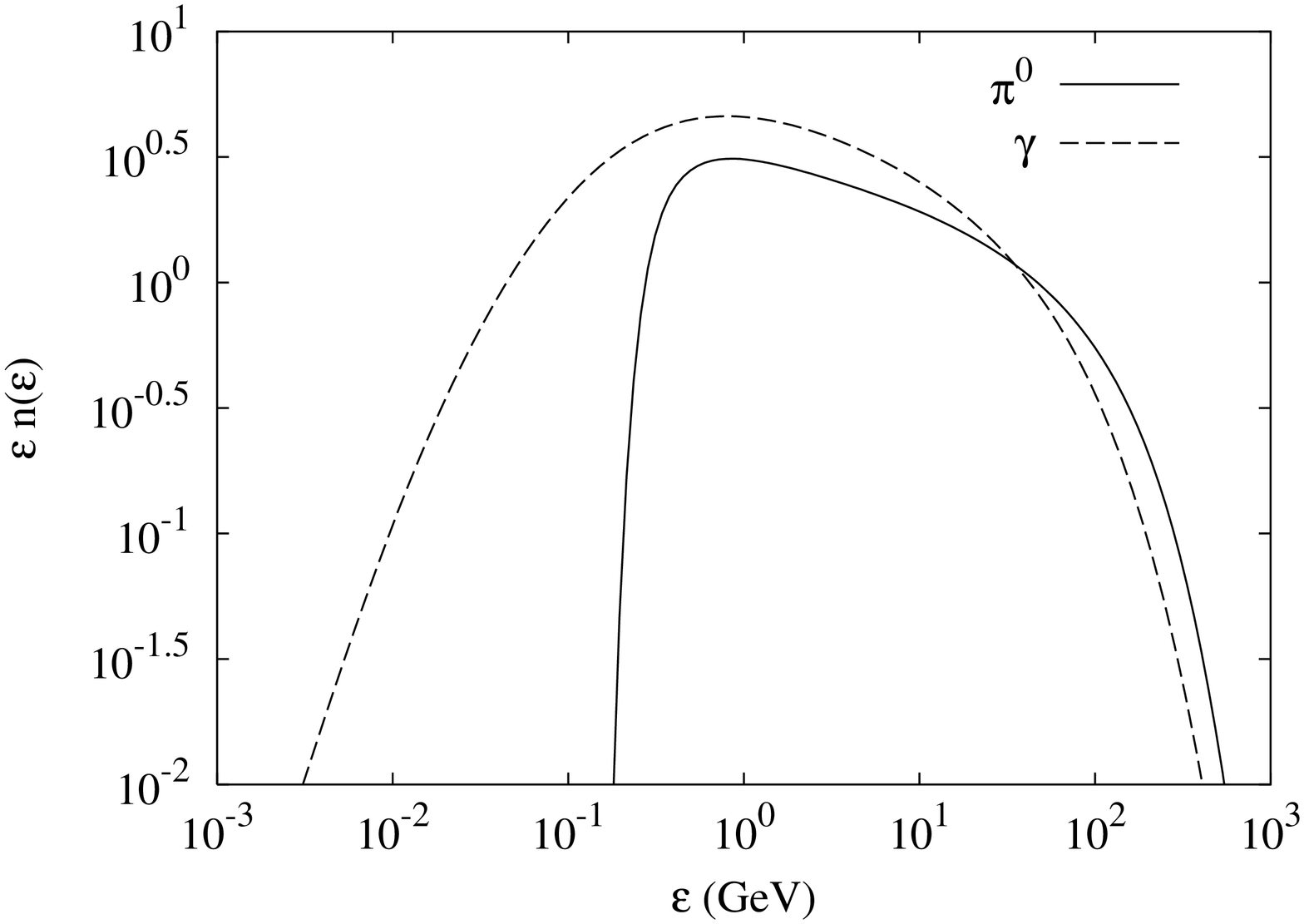}
\end{center}
\caption{Energy spectra of $\pi^0$ mesons created in
a $pp$ collision and of the resulting gamma rays.
Top panel: lab frame, corresponding to a proton with energy $E_p=10^6$ GeV colliding
with a proton at rest; bottom panel: center-of-mass frame, corresponding to two protons
with energy 730 GeV colliding head-on.}
\label{fig:app:pizero}
\end{figure}

First, we consider the energy and rapidity distribution
of $\pi^0$  mesons created in a collision of an energetic proton $p$
with a fixed-target proton $q$. This is the scattering geometry for which
the parameterizations presented in this work are derived. We denote
the Lorentz frame corresponding to this scattering geometry with $K$ throughout this section. In figure \ref{fig:app:lab}, we present the 
$\pi^0$ distribution for incident proton energy $E_p = 10^6$ GeV. We observe that
the energy and rapidity of the secondary pions are strongly correlated: pions with higher energies are emitted closer to the direction of the incoming
energetic proton (corresponding to higher values of the rapidity $y$;
see eq. \eqref{kin:defrap}).

Next, we consider two protons $p$ and $q$ colliding head-on with energies $E_p'$ and $E_q'$ which defines the reference frame $K'$. Without loss of generality, we
take the protons  to be moving in the $z'$ direction.
The secondary particle distribution in this frame is given by
\beqs
\label{eq:kin:Lorentz}
n'(\ep',y') = \lb \frac{\ep'}{\ep} \frac{\cosh^2 y}{\cosh^2 y'} \rb n(\ep,y) \, ,
\eeqs
which follows from 
eq. \eqref{eq:theory:n} and the invariance of $\ep \, n(p_x,p_y,p_z)$
In this equation, $n(\ep,y)$ is the particle distribution in the frame $K$ which is parameterized
in eq. \eqref{eq:res:dist:pion}.
Note that eq. \eqref{eq:kin:Lorentz} is only valid if the frames $K$ and
$K'$ are connected with a single Lorentz boost in the $z$ ($z'$) direction,
i.e. for protons colliding head-on along the $z'$ axis in the $K'$ frame.

As a concrete example, we consider two protons with equal energies
$E_p' = E_q' = 730$ GeV. In this case, $K'$ coincides with
the center-of-mass (COM) frame for a collision between a 
proton with energy $E_p =10^6$ GeV and a proton at rest. In particular, this means that
the center-of-mass energy $\sqrt{s}$ and the secondary multiplicities
are identical for the scattering geometries in the frames $K$ and $K'$. 

In figure \ref{fig:app:com}, we show the $\pi^0$ energy and rapidity 
distribution after the collision in the COM frame $K'$.
In this frame, the scattering geometry is invariant under the
interchange of the two protons so that the
secondary particle distribution is symmetric under the transformation
$y \to -y$. It is observed from the figure that this is indeed the case
for the distribution derived from the parameterization presented in this
work. This is an  
\emph{a posteriori} verification of the parameterization, which is derived
in the lab frame without considering this symmetry.

\subsubsection{Energy spectrum of secondary particles and decay products}
In figure \ref{fig:app:pizero}, we show the secondary $\pi^0$ energy spectra 
for the scattering geometries associated with the $K$ and $K'$ frames,
together with the gamma-ray energy spectra resulting from the decay
$\pi^0  \to \gamma \gamma$. The decay spectrum $n_\gamma ( \ep_\gamma)$ is 
related to the pion spectrum $n (\ep)$ as follows (see, e.g., Ref. \cite{1981Ap&SS..76..213S}):
\beqs
n_\gamma (\ep_\gamma) = 2 \int^{\infty}_{\ep_\gamma + m_\pi^2 c^4/ 4 \ep_\gamma}
\frac{n(\ep)}{\sqrt{\ep^2 - m_\pi^2 c^4}} \, d \ep \, ,
\eeqs
where $n (\ep)$ is the $\pi^0$ energy spectrum. Because this formula
is valid in all frames, $n$ and $\ep$ may be replaced by $n'$ and $\ep'$ to derive
the gamma-ray energy spectrum from the pion energy
spectrum in the $K'$ frame.

\subsection{Proton -- proton collision at an arbitrary angle}
\label{sec:appl:genang}
In this section, we  consider  two protons with
energies $E_p'$ and $E_q'$ that collide at an arbitrary angle.
Without loss of generality, we take proton $p$ to be moving along the $x'$ axis in the
$+x'$ direction and proton $q$ to be moving in the 
$x'$ -- $y'$ plane at an angle $\phi_p'$ with respect to the $x'$ axis.

We parameterize the distribution of secondary $\pi^0$ mesons created
in this interaction with
the pion energy $\ep'$, the zenith
angle $\theta_\pi'$ (with respect to the $z'$ axis) and the 
azimuthal angle $\phi_\pi'$ (in the $x'$ -- $y'$ plane).
The pion momentum is thus expressed as follows:
\bsub
\label{eq:def:pionangles}
\beqs
k'_x &=& |\vec{k}'| \sin \theta_\pi' \cos \phi_\pi'  \, ; \\
k'_y &=& |\vec{k}'| \sin \theta_\pi' \sin \phi_\pi' \, ; \\
k'_z &=& |\vec{k}'| \cos \theta_\pi'  \, ,
\eeqs
\esub
where  $c |\vec{k}'|  = \sqrt{\ep'^2 - m_{\pi}^2 c^4}$.
In the following, we derive the secondary $\pi^0$
angular distribution in the scattering plane
and the $\pi^0$ energy spectrum.

\subsubsection{Secondary angular distribution in the scattering plane}
\begin{figure}
\begin{center}
\includegraphics[width=6cm]{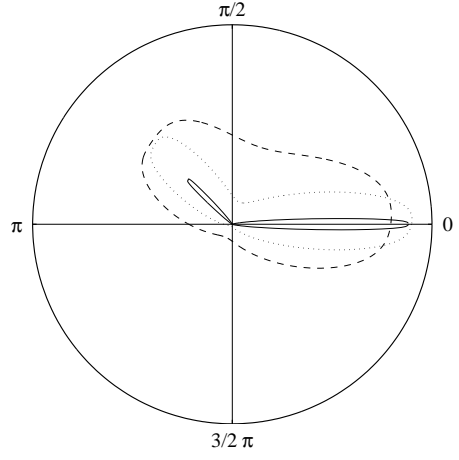}
\end{center}
\caption{Polar plot of the $\pi^0$ distribution $ n'(\ep',\phi_\pi')$
as a function of the azimuth angle $\phi_\pi'$
after a collision of a $10^4$ GeV proton with a
$10^2$ GeV proton at an angle $\phi_q' = (3/4) \pi$.
We plotted the distribution  
for pion energies $\ep'=5$ GeV (solid line), $\ep'=1$ GeV (dotted line) and $\ep'=0.5$ GeV (dashed line).}
\label{fig:app:corr1}
\end{figure}

\begin{figure}
\begin{center}
\includegraphics[angle=270,width=8.6cm]{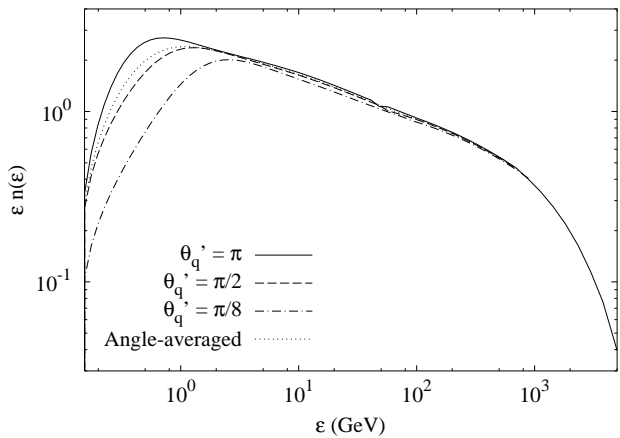}
\end{center}
\caption{Energy spectra of secondary $\pi^0$ mesons created in a collision
of a $10^4$ GeV proton with a $10^2$ GeV proton for three different incident angles
$\theta_q'$. Also shown is the angle-averaged spectrum (see text).
For numerical reasons we only plot the energy spectrum 
for head-on collisions at energies \mbox{$\ep \gtrsim 10^3$ GeV}.
We have verified with $\texttt{PYTHIA}$ simulations that this part
of the spectrum is independent of the incident angle between the protons.}
\label{fig:app:spec}
\end{figure}

\begin{figure}
\begin{center}
\includegraphics[width=6.8cm]{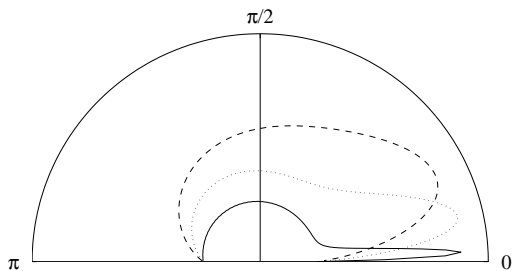}
\end{center}
\caption{Polar plot of the $\pi^0$ distribution $n'(\ep',\theta_\pi')$
as a function of the zenith angle $\theta_\pi'$ after a collision
of a $10^4$ GeV proton with an isotropic distribution of $10^2$ GeV protons.
The meaning of the lines is the same as in figure
\ref{fig:app:corr1}.}
\label{fig:app:corr2}
\end{figure}

The pion distribution in the frame $K'$ is derived from the parameterization in the fixed-target frame $K$ by Lorentz transformations.
The frames $K'$ and $K$ are connected 
by a Lorentz boost
to the rest-frame of proton $p$, followed by a rotation to align the incoming
proton $q$ with the $z$ axis. The number of
secondary pions with  energy and angles in the range
$(\ep' \ldots \ep' + d \ep') \times
(\theta'_\pi \ldots \theta'_\pi + d \theta'_\pi) \times
(\phi'_\pi \ldots \phi'_\pi + d \phi'_\pi)$ is equal to 
$ n'(\ep',\theta'_\pi,\phi'_\pi) \sin \theta'_\pi  d \theta'_\pi
d \phi'_\pi d\ep'$, where
\beqs
\label{eq:dist:usepar}
n'(\ep',\theta'_\pi,\phi'_\pi)
=  \lb \frac{\ep \sqrt{\ep'^2 - m_{\pi}^2 c^4}}{m_{\pi}^2 c^4 + \ep^2 (1 - \tanh^2 y)} \rb  \frac{n(\ep,y)}{2 \pi}  \, .
\eeqs
In this formula, $\ep$ and $y$ are the pion energy and rapidity in the $K$ frame, respectively, and $ n(\ep,y)$ denotes the pion energy and rapidity distribution which is parameterized in eq. \eqref{eq:res:dist:pion}.

In figure \ref{fig:app:corr1}, we present the distribution of secondary $\pi^0$ mesons
with respect to the azimuthal angle
$\phi_\pi'$, i.e.,
\beqs
n'(\ep',\phi'_\pi) \equiv \int_{0}^\pi n'(\ep',\theta'_\pi,\phi'_\pi) \sin \theta'_\pi d \theta'_\pi \, ,
\eeqs
for different values of the pion energy $\ep'$. In producing this figure, we have chosen incident proton energies $E_p' = 10^4$ GeV and $E_q' = 10^2$ GeV and incident angle $\phi_q'= (3/4) \pi$. As can be
seen from the figure, the pions are produced mostly in the direction of the
incident protons. The degree of collimation is correlated with the energy:
for pion energies $\ep'$ below a few GeV, where the pion spectrum is
highest (see fig. \ref{fig:app:spec}), the pion direction 
can be significantly different from the direction of the colliding protons.
At energies above a few GeV, the angle of the outgoing pion
is typically within a few degrees of the direction of one of the colliding
protons. 
We have verified that this result holds for all secondary pions and kaons.

\subsubsection{Secondary energy spectrum}
In figure \ref{fig:app:spec} we present 
pion energy spectra (integrated over pion angles)
resulting from a collision of two protons
with energies $E_p' = 10^4$ GeV and $E'_q = 10^2$ GeV for different
values of the proton collision angle $\phi_q'$. 
For comparison, we also show in this figure the pion spectrum
averaged over incident proton angles (see below).

While the energy spectrum at high energies is independent of the
incident proton angle, there are significant differences at
low energies. 
For  small values of the incident proton angle $\phi_q'$, i.e.
close to a tail-on collision, the low-energy part of
the spectrum is suppressed as expected.

\subsection{Isotropic distribution of target protons}

In this section, we consider a single high-energy proton $p$ with energy 
$E_p'$ that interacts with an isotropic distribution (in three dimensions)
of  mono-energetic low-energy protons $q$ with energy $E_q'$. 
We derive the distribution of secondary pions with respect to the angle
between the high-energy proton and the pion, as well as the
energy spectrum. For an 
isotropic distribution of target protons, the resulting
pion distribution does not depend on the azimuthal angle
around the direction of the high-energy proton. In order
to keep the former definition of pion angles (eqs. 
\eqref{eq:def:pionangles}) we consider in this section
a high-energy incident proton that moves
along the $z'$ axis in the $+z'$ direction.
With this choice, the zenith angle between the high-energy proton and the pion
is equal to $\theta'_\pi$.

The momentum of proton $q$ is
expressed in terms of angles in the same way as the pion
momentum in eqs. \eqref{eq:def:pionangles}: the angle
$\theta'_q$ denotes
the zenith angle with respect to the $z'$-axis 
and $\phi'_q$ denotes the azimuth angle with respect to the
$x'$ axis in the $x'$ -- $y'$ plane.

\subsubsection{Zenith angle distribution of secondary pions}
The secondary pion distribution, averaged over the incoming angles of low-energy protons $q$,
is given by the following expression:
\beqs
\nn \bar{n}'(\ep',\theta'_\pi,\phi'_\pi) & \equiv & 
\frac{d^3 \bar{N}}{d \ep'  d \cos \theta_\pi'  d \phi_\pi'  } \\
\label{eq:def:avn} & = &  \frac{1}{\bar{\sigma}'_{\textrm{inel}}}
\frac{d^3 \bar{\sigma}'}{d \ep'  d \cos \theta_\pi'  d \phi_\pi'  } \, , 
\eeqs
where $N$ is the total number of created pions,
$\sigma'_{\textrm{inel}}$ is the inelastic $pp$ cross section
and $\sigma$ is the inclusive cross section to detect a
particle of a given species assuming an ideal detector
(cf. section \ref{sec:dist}). In this section, we use a bar to
indicate that a quantity
is averaged over 
the incoming angles of low-energy protons $q$.

For clarity we assume in this section that both protons are very energetic,
so that we may take $\beta'_p = \beta'_q = 1$. The averaged
inelastic cross section is then equal to
\beqs
\label{eq:def:avsinel}
\bar{\sigma}_{\textrm{inel}}' = \frac{1}{2} \int_0^\pi d \theta_q'
\sin \theta_q' (1 - \cos \theta_q') \, \sigma_{\textrm{inel}} (s(\theta_q')) \, .
\eeqs
In this equation, $\sigma_{\textrm{inel}}$ depends
on the proton angle $\theta'_q$ through 
the center-of-mass energy $\sqrt{s}$, where
\beqs
s(\theta'_q) = 2 m_p c^2 + 2 E'_p E'_q (1 - \cos \theta'_q) \, .
\eeqs
The dependence of the inelastic cross section on $s$ is expressed in eqs.
\eqref{eq:fitsigmainel} and \eqref{eq:Ep:sqrts}.

For given values of the proton angles $\theta'_q$
and $\phi'_q$, the differential inclusive cross section and
the secondary particle distribution are related as follows:
\beqs
\left.
\nn \frac{d^3 \sigma'}{d \ep'  d \cos \theta'_\pi d\phi'_\pi } 
\right|_{\theta'_q, \phi'_q,} \hspace{-0.1cm} & = & \hspace{-0.1cm}  (1 - \cos \theta_q')  \sigma'_{\textrm{inel}} (s (\theta_q')) \\ 
\label{eq:def:dsigmaincl} & & \hspace{-0.1cm}   \times  \, n' ( \ep',  \theta'_\pi, \phi'_\pi; \theta'_q, \phi'_q)  \, ,
\eeqs
where we have explicitly written the dependence of the pion
distribution $n'$ on the proton angles $\theta_q'$ and $\phi_q'$.
The total inclusive cross section $\bar{\sigma}'$ is obtained
by integrating eq. \eqref{eq:def:dsigmaincl}
over the outgoing pion angles and
averaging over the incident proton angles.
The resulting pion distribution is homogeneous in the $\phi'_\pi$ variable.
We use this rotational invariance
to replace the integral over $\phi_q'$ with a factor $2 \pi$ and choose
the value  $\phi'_q =0$ to find:
\beqs
\nn \bar{\sigma}' \hspace{-0.1cm} &= & \hspace{-0.1cm} \frac{1}{2}\int d \ep' d \theta'_\pi d \phi'_\pi \sin \theta'_\pi \int d \theta'_q \, \sin \theta'_q (1 - \cos \theta_q') \\
\label{eq:def:avsincl} &  & \hspace{-0.1cm} \times \,  \sigma'_{\textrm{inel}} (s (\theta_q')) \, 
n' ( \ep', \theta'_\pi, \phi'_\pi; \theta'_q,  \phi'_q = 0) \, ,
\eeqs
where the integrals cover the full phase space.
The pion distribution with respect to the 
pion energy $\ep'$ and scattering angle $\theta'_\pi$ is defined as:
\beqs
\bar{n}'(\ep',\theta'_\pi) = \frac{d^2 \bar{N}}{d \ep' d \theta'_\pi}
= \sin{\theta'_\pi} \int_0^{2 \pi} d \phi'_\pi \frac{d^3 \bar{N}}{d \ep' d \cos \theta'_\pi d \phi'_\pi}  \, .
\eeqs
Using eqs. \eqref{eq:def:avn} and
\eqref{eq:def:avsincl}, we find that
\beqs
\nn \bar{n}'(\ep',\theta'_\pi) \hspace{-0.1cm} &= & \hspace{-0.1cm}
\frac{\sin \theta'_\pi}{2 \, \bar{\sigma}_{\textrm{inel}}'} 
 \int_0^\pi d \theta'_q \, \sin \theta'_q 
(1 - \cos \theta_q')  \sigma'_{\textrm{inel}} (s (\theta_q'))  \\
\label{eq:def:avnth} & & \hspace{-0.1cm} \times \, 
\int_0^{2 \pi} d \phi'_\pi \, 
n' ( \ep', \theta'_\pi, \phi'_\pi; \theta'_q,  \phi'_q = 0) \, ,
\eeqs
where $\bar{\sigma}_{\textrm{inel}}'$ is defined in eq. \eqref{eq:def:avsinel}.

In figure \ref{fig:app:corr2} we show the
distribution $\bar{n}' (\ep',\theta'_\pi)$ as a function of $\theta'_\pi$ for
three different values of the pion energy $\ep'$. In producing this figure,
we have considered a collision
of an energetic proton with energy $E_p' = 10^4$ GeV with an isotropic
distribution of mono-energetic protons with energy $E_q' = 10^2 $ GeV.
From the figure we observe that
pions with higher energy are collimated stronger within the direction 
of the incoming proton, as expected. We have verified that this
holds for all secondary mesons.
In the maximally forward direction,
i.e., near $\theta'_\pi = 0$, the distribution decreases
because the available phase space is proportional to $\sin \theta'_\pi$.

\subsubsection{Energy spectrum of secondary pions}
The secondary $\pi^0$ energy spectrum for the interaction of
a $10^4$ GeV proton with the distribution of
$10^2$ GeV protons is expressed as
\beqs
\bar{n}'(\ep') = \int_0^\pi d \theta'_\pi \, \bar{n}'(\ep',\theta'_\pi) \, ,
\eeqs
where $\bar{n}'(\ep',\theta'_\pi)$ is given in eq. \eqref{eq:def:avnth}.
We show in figure  \ref{fig:app:spec} the energy spectrum averaged over
the incoming angles of the low-energy protons $q$.
We find that the averaged spectrum 
is very close to the spectrum resulting from a collision of
a $10^4$ GeV proton with a $10^2$ GeV proton
with incident angle $\theta'_q \simeq (5/8) \pi$, i.e. in the forward direction but not
head-on. Qualitatively, this is as expected because
the cross section $\sigma'_{\textrm{inel}} (s (\theta_q')) $ and the flux factor
$(1 - \cos \theta_q')$ are largest for head-on collisions while the
phase-space volume factor $\sin \theta_q'$ suppresses head-on collisions.

\section{Extrapolation to the highest cosmic-ray energies}
\label{sect:disc}
\begin{figure}
\begin{center}
\includegraphics[angle=270, width=8.6cm]{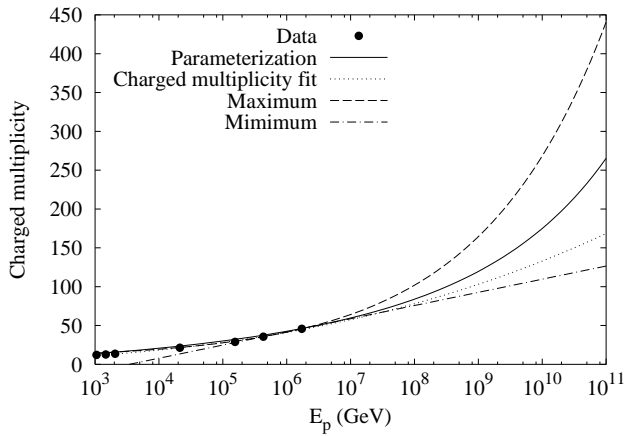}
\end{center}
\caption{Charged multiplicity as a function of incident proton energy.
The solid line shows an extrapolation of eq.
\eqref{eq:theory:chmult}; the dotted line shows the multiplicity estimated from the
parameterized charged pion distributions presented in this work; the 
dashed and dash-dotted lines show the minimum and maximum multiplicities 
given in eqs. \eqref{mult:minmax}. The data is taken from Refs. \cite{Ansorge:1988kn, Alpgard:1982kx,Benecke:1974if, Biyajima:2001ud, Breakstone:1983ns}.}
\label{fig:disc:mult}
\end{figure}
The parameterizations presented in section \ref{sec:results} are based on
simulated $pp$ collisions for
incident proton energies \mbox{$10^3 \textrm{ GeV} < E_p < 10^6 \textrm{ GeV}$}, where
experimental data is available to verify the resulting
particle distributions and multiplicities.
Cosmic-ray observations suggest that the maximum proton energy that can be generated in astrophysical proton accelerators may be as high as  $10^{11} \textrm{ GeV}$.
Thus, in order to account for interactions of the highest energy protons,
the parameterizations presented in this
work need to be applied in a region where they cannot be directly tested.
In this section, we compare the high-energy  behavior of the parameterizations
derived in this work with  an extrapolation of existing data
and with theoretical models.
Extrapolations of 
experimental data as well as theoretical models for
incident proton energies $ E_p > 10^6 \textrm{ GeV}$
are available predominantly for the charged multiplicity, due to the availability
of experimental data at lower energies.
Therefore, we focus in this section on
the charged multiplicity contained in the parameterizations
presented in this work.

The charged multiplicity is dominated by pions, hence we estimate the
charged multiplicity from the parameterized charged pion distributions.
We derive the
charged pion multiplicity  $\mathcal{M}^\textrm{par}_{\pi^\pm} =  \mathcal{M}^\textrm{par}_{\pi^+}
+ \mathcal{M}^\textrm{par}_{\pi^-}$ by 
integrating eq. \eqref{eq:res:dist:pion} over
energy and rapidity. To account for charged particle creation due to decay processes 
and for the contribution of other charged particles,
we estimate the charged multiplicity with
$ \mathcal{M}^\textrm{par}_{\textrm{ch}} = 2 + 1.47 \, \mathcal{M}^\textrm{par}_{\pi^\pm} $, where the numerical value $1.47$ is found by comparing $\mathcal{M}^\textrm{par}_{\pi^\pm}$ and $\mathcal{M}_{\textrm{ch}}$ at the proton energies considered in our simulations. The leading term $2$ accounts for the number of outgoing protons
for low secondary multiplicities (corresponding to low center-of-mass energies).

Using experimental data at low energies,  \citet{Engel:1998hf} has found that the 
charged multiplicity should increase faster than $\log (s)$ but not as fast as $s^p$, where \mbox{$0.1 < p <  0.3$}, at high energies. In order to compare our results
with these limiting cases, we have re-derived\footnote{The explicit
form of these functions was not given in the original work 
\cite{Engel:1998hf}. The numerical value of 0.22 is chosen for comparison with fig. 8 of Ref. \cite{Engel:1998hf}.} the 
explicit functional form based on the two data points with highest
energy \cite{Ansorge:1988kn,Biyajima:2001ud}:
\bsub
\label{mult:minmax}
\beqs
\mathcal{M}^\textrm{min}_{\textrm{ch}} & = & -65 + 17 \log s \, ; \\
\mathcal{M}^\textrm{max}_{\textrm{ch}} & = & 7.0 + 1.4 \, s^{0.22} \, ,
\eeqs
\esub
where $s$ is expressed in units of GeV$^2$.
In figure \ref{fig:disc:mult} we show the charged multiplicity estimated
from the parameterizations presented in this work, together with the minimum and maximum
values of the multiplicity given in eqs. \eqref{mult:minmax}.
Also shown is an extrapolation of the approximation $\mathcal{M}^\textrm{fit}_{\textrm{ch}}$ given in eq. \eqref{eq:theory:chmult}.
We observe that the charged multiplicity estimated from our parameterizations 
increases faster than the extrapolation of $\mathcal{M}^\textrm{fit}_{\textrm{ch}}$
but is well within the limits derived by \citet{Engel:1998hf}.
We thus conclude that the high-energy behavior of the parameterizations
presented in this work is consistent with theoretical expectations.

Applying the parameterizations presented in this work to very high
proton energies $E_p > 10^6$ GeV should be done with caution because there is no
experimental data to directly test the results. Nevertheless,
we estimate from figure \ref{fig:disc:mult}
that the uncertainty in the charged particle multiplicity at the highest energies 
$E_p \simeq 10^{11}$ GeV is less than a factor 2.
While lacking experimental data on the full particle distributions at high
energies, we find a strong similarity 
in the particle distributions for different  proton energies  within
the three orders of magnitude in energy range considered in this work.
This suggests
that the shape of the distributions does not change significantly
at very high energies so that the parameterizations derived in this work
can be applied with some confidence
to $pp$ collisions with proton energies $E_p > 10^6$~GeV.

\section{Discussion}
\label{sec:discussion}

\subsection{Comparison with previous work}
\label{sec:disc:val}
\begin{figure}
\begin{center}
\includegraphics[angle=270, width=8.6cm]{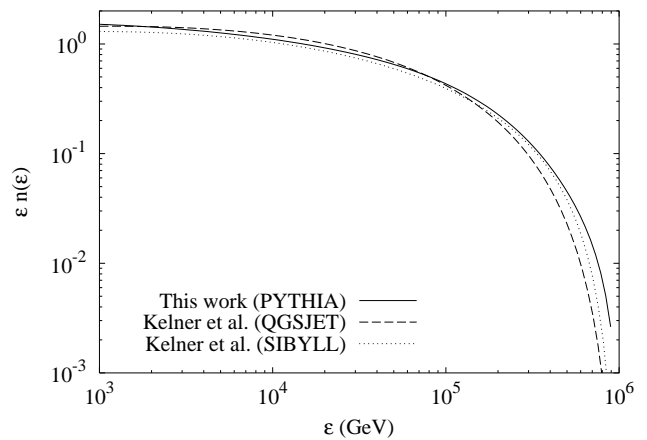}
\end{center}
\caption{Comparison of parameterizations of the $\pi^0$ 
energy spectrum $\ep \, n(\ep)$ for
incident proton energy $E_p=10^6$ GeV. The energy range is chosen for
comparison with fig. 1 of  Ref. \cite{Kelner:2006tc}.}
\label{fig:res:comp}
\end{figure}
We have verified that
the parameterizations of the pion and charged kaon distributions presented in this work are similar to those by
\citet{Badhwar:1977zf} and \citet{1981Ap&SS..76..213S} at the lowest energies considered
in this work, $\sqrt{s} = 45$ GeV.
The difference between our parameterization of the neutral pion distribution and that
of \citet{Blattnig:2000zf} is larger. These authors provide an accurate fit
to the particle distribution at large transverse momentum. However, the number of
particles in this region is very small and we find that the parameterization does
not reproduce the total multiplicity correctly for center-of-mass energy
\mbox{$\sqrt{s} = 45$ GeV}.

In figure \ref{fig:res:comp}, we present a comparison of the
parameterization of the $\pi^0$ energy spectrum presented in this work
with two parameterizations by \citet{Kelner:2006tc}.
These parameterizations  are based on Monte Carlo results generated with \texttt{QGSJET}
and \texttt{SYBILL} instead
of \texttt{PYTHIA}. 
We observe from the figure that our parameterization is closer to 
the  \texttt{QGSJET} fit at intermediate energies
and closer to the \texttt{SYBILL} fit at high energies.
The differences between the energy spectra described by 
the three parameterizations are up to $\sim$$10$\% for intermediate pion energies.
This is larger than the fit inaccuracy (see section \ref{sec:results}), which
suggests that a more precise description of the energy spectra and
particle distributions requires a better theoretical understanding of 
the $pp$ physics, in particular of the fragmentation process,
rather than more accurate fits.

The incomplete understanding of the
collision physics is reflected in the availability of
different tuning models for \texttt{PYTHIA}. 
Several tuning models for \texttt{PYTHIA} version 6.2 can be found in the literature,
such as CDF tune A (Field\footnote{The CDF tune A is available
at the website \texttt{http://www.phys.ufl.edu/\~{ }rfield/cdf/tunes/py\_tuneA.html}.};
see also Ref. \cite{Buttar:2005}) or the tuning
proposed in Ref. \cite{Buttar:2004iy}.
In this work, we have used  \texttt{PYTHIA} 6.3 with default values. A
preliminary tuning for \texttt{PYTHIA} version 6.3 is presented in Ref.
\cite{Moraes:2006}. It is our intention to adopt such a tuning after it
has been extensively tested against the available data. In this work, we have
not studied the effect of changing \texttt{PYTHIA} parameters, but we expect
the differences in the resulting particle distributions to be comparable to
the those resulting from the use of different Monte Carlo event generators
(see fig. \ref{fig:res:comp}).

\subsection{Astrophysical applications}
\label{sec:disc:app}
In section  \ref{sec:app} we demonstrated through explicit examples
how the parameterizations presented in this work can be used to study particle creation in
collisions of protons with different energies and an arbitrary incident angle.
We found that the energy of secondary $\pi^0$ mesons
and the degree of collimation around the direction of the colliding protons
are strongly correlated (see figures \ref{fig:app:corr1} and \ref{fig:app:corr2}).
We have verified that this holds for all secondary mesons. We expect that this
has interesting observational consequences for neutrinos and gamma rays created in
$pp$ collisions in astrophysical sources.

In the early prompt emission of GRBs, 
the optical depth for $pp$ interactions can be larger than a few.
The mechanism responsible for the dissipation of 
the fireball energy may accelerate a substantial fraction of the 
protons contained in the fireball to high energies. Subsequent
$pp$ collisions give rise to high-energy neutrinos and gamma rays.
Because the high-energy
secondary mesons are collimated within the direction of the energetic 
proton, the energy and angular distribution of these secondaries
depends on the distribution of the high-energy protons. 
Therefore, the resulting neutrino and gamma-ray signals may contain
information about the details of the acceleration mechanism.
We note however that in this
scenario both interacting protons are moving toward an observer with ultra-relativistic
velocities so that all particle distributions are collimated by relativistic
beaming. This implies that also low-energy secondaries will be collimated in
the observer frame and it will be difficult to extract information from
the resulting signals. Nevertheless, the angle-energy correlations found
in this work may make it possible to extract information about the acceleration
mechanism.

The effect of energy-dependent collimation is particularly important for 
observations of secondary particles produced in $pp$
collisions where the target proton is non-relativistic in the
observer frame. A very interesting scenario in this respect
is that of `failed GRBs' \cite{Meszaros:2001ms,Razzaque:2003uv} (see also
Refs. \cite{Razzaque:2004yv,Razzaque:2005bh}).
It is possible that the mechanisms associated with the early
phases of a developing GRB may be present in a large fraction of supernovae,
but only lead to an observed GRB under special circumstances. For example, 
it may be the case that the formation
of a fireball is quite a common phenomenon but that a large fraction
of fireballs has insufficient energy to traverse the pre-burst
stellar environment. However, if shocks form at a sub-stellar radius,
protons can be accelerated and collide with target protons (nuclei).
The resulting neutrinos are likely the only signals that could
indicate the existence of such a class of failed (dark) GRBs. In this scenario,
the energy and flux of the neutrinos that reach the earth depend
strongly on the collimation of the secondary neutrinos created in the $pp$
collisions.

\section{Conclusion}
\label{sec:conclusion}

In this paper we have studied the creation of secondary pions and kaons in energetic
$pp$ collisions using the event generator \texttt{PYTHIA}.
We considered an incident proton with energy \mbox{$10^3 \textrm{ GeV} < E_p  < 10^6 \textrm{ GeV}$}
colliding with a target proton at rest. This corresponds to
center-of-mass energy \mbox{$43 \textrm{ GeV} < \sqrt{s} < \EE{1.4}{3} \textrm{ GeV}$}.
The key result is the parameterization of the 
energy spectra  (eq.  \eqref{eq:spec:res}) and the
energy and rapidity distributions 
(eqs. \eqref{eq:res:dist:pion}  and \eqref{eq:res:dist:kaon})
of secondary pions and kaons.
Applications of these parameterizations are presented in section \ref{sec:app}.
In section \ref{sect:disc}, we have argued that
the results can be applied to $pp$ interactions for protons with energies
$E_p >  10^{6}$~GeV. 
At the highest CR energies, $E_p \simeq 10^{11}$~GeV, we
have estimated the uncertainty in the parameterizations to be within a factor 2.

We have parameterized the particle distributions of meta-stable pions and kaons, as opposed to
stable decay products, because this captures the essential properties of the $pp$ interaction
without making any assumptions about the importance of pion and kaon energy loss
prior to decay (for concrete implications of pion decay in an astrophysical context,
see e.g. Refs. \cite{Kashti:2005qa,Asano:2006zz}).
Energy spectra and full particle distributions of neutrinos and gamma rays
are derived from the parameterizations presented in this work in a 
straightforward manner.

The energy and rapidity distributions fully describe the kinematics of 
the secondary mesons, so that
the derived parameterizations contain all correlations between energy and angle
of the outgoing particles and can be applied to a general scattering geometry,
viz. two protons with different energies that collide under an arbitrary
angle. Hence the parameterizations are 
not limited to collisions where one proton is at rest, which
opens a wealth of astrophysical applications.

We presented in section \ref{sec:app} several applications
of the derived parameterizations. We derived the gamma-ray spectrum resulting
from $\pi^0$ decay after a $pp$ collision (see figure \ref{fig:app:pizero}) 
and we studied correlations between the energy and outgoing angle of secondary $\pi^0$ mesons produced
in a collision of two protons with different energies and arbitrary incident angle
(see figure \ref{fig:app:corr1}).
These results can be used for a detailed study of $pp$ interactions in the early
prompt emission of GRBs and in the interaction of a developing GRB with its surroundings
(see section \ref{sec:disc:app}).
A particularly interesting possibility is the existence of a class of developing GRBs
for which the fireball has insufficient energy to traverse the pre-burst stellar environment.
If shocks are formed at a sub-stellar radius, these will accelerate protons
that collide with target protons and create neutrinos.
The fluence and energy of neutrinos that reach the earth 
depend sensitively on the correlations between the energy and outgoing angle of the 
secondary mesons that are created in these $pp$ interactions.
The parameterizations presented in this work can be used to study this
scenario in detail. We aim to investigate this in the future.

The parameterizations presented in this work can be extended with proton -- neutron
and proton -- photon interactions,  all of which can be studied with 
\texttt{PYTHIA}.  The same holds for the
energy spectrum and angular distribution
of primary nucleons (the primary nucleon is the outgoing nucleon with the highest energy).
This allows a more precise study of multiple nucleon -- nucleon interactions.
These parameterizations are left for future work.

\begin{acknowledgments}
We wish to thank James Miller-Jones and Peter M\'esz\'aros for useful discussions. 
H.~K. wishes to thank Justus Koch and Paul van der Nat for valuable discussions on hadron physics.
This research was supported by NWO grant 639.043.302 to R.W. and by the
EU under RTN grant HPRN-CT-2002-00294.
\end{acknowledgments}

\appendix

\section{The Lund string model}
\label{sec:applund}
The Lund string model \cite{Andersson:1983ia} is an iterative model
used in \texttt{PYTHIA} to describe meson formation after a hard QCD process.
In the model, quark-antiquark pairs that are created in a hard
QCD scattering process form `strings' 
that are connected through a color flux-tube with energy per unit length $\kappa$. 
This string breaks 
into a meson and a remainder
string that will undergo the same process (baryons are generated
through a generalization of this process).
At every step in the iteration, a meson is created with a certain energy and rapidity 
according to a probability distribution. 

The mechanism to break the string is the creation of a  new quark-antiquark pair
through quantum-mechanical tunneling. The probability to
create a $q \bar{q}$ pair with mass $m$ and transverse momentum $p_T$ is given by 
\beqs
\label{eq:mult:Schwinger}
\mathcal{P} = \exp \lb - \frac{\pi}{\kappa} \lb m^2 c^4 + p_T^2 c^2\rb \rb \, ,
\eeqs
which derives from the Schwinger formula \cite{Schwinger:1951nm}. 
This implies that lighter mesons are created more prolifically than heavier
mesons and that the probability to create a meson falls off exponentially
with increasing $p_T$.
After a meson is created, the probability
that it carries a fraction $z$ of the string's $E+p_z$ is determined
by the so-called 
fragmentation function \cite{Andersson:1983ia,Sjostrand:2003wg}.
Together with eq. \eqref{eq:mult:Schwinger}, this fragmentation function determines the secondary 
particle distributions. Free parameters within the model are adjusted to reproduce
experimental data. A detailed description can be found in Refs. \cite{Andersson:1983ia,Sjostrand:2003wg}.

\begin{table*}
\caption{\label{table:num:spec} Numerical values of the energy spectrum fit parameters $p_{ij}$.}
\begin{ruledtabular}
\begin{tabular}{c c c c c c c c }
 & $\pi^+ $& $\pi^-$ & $\pi^0$ &  $K^+ $& $K^-$ & $K^0$ & $\overline{K^0} $  \\
\hline
$ p_{00}  $ & $  -0.497  $ & $  -0.501  $ & $ -0.456   $ & $ -1.23  $ & $  -1.46  $ & $  -1.29  $ & $  -1.50  $ \\ 
$ p_{01}  $ & $  0.0934  $ & $  0.0934  $ & $ 0.0950   $ & $ 0.0657  $ & $  0.101  $ & $  0.0690  $ & $  0.101  $ \\ 
$ p_{10}  $ & $  -0.140  $ & $  -0.128  $ & $ -0.142   $ & $ -0.147  $ & $  -0.109  $ & $  -0.142  $ & $  -0.118  $ \\ 
$ p_{11} $ & $  0.0131  $ & $  0.0118  $ & $ 0.0135   $ & $ 0.0161  $ & $  0.00865  $ & $  0.0154  $ & $  0.0101  $ \\ 
$ p_{20}  $ & $  -0.455  $ & $  -0.437  $ & $ -0.457   $ & $ -0.00411  $ & $ -0.0577  $ & $  -0.00717  $ & $ -0.0567  $ \\ 
$ p_{21} $ & $  0.495  $ & $  0.494  $ & $ 0.494   $ & $ 0.493  $ & $  0.491  $ & $  0.493  $ & $  0.491  $ \\ 
$ p_{30}   $ & $  -2.06  $ & $  -0.945  $ & $ -1.49   $ & $ -0.989  $ & $  -1.22  $ & $  -1.03  $ & $  -1.25  $ \\ 
$ p_{40} $ & $  -0.896  $ & $  -1.03  $ & $ -0.981   $ & $ -0.345  $ & $  -0.164  $ & $  -0.294  $ & $  -0.169  $ \\ 
$ p_{50}   $ & $  1.11  $ & $  0.963  $ & $ 1.01   $ & $ 0.777  $ & $  1.04  $ & $  0.839  $ & $  1.05  $ \\ 
$ p_{60}$ & $  0.791  $ & $  0.598  $ & $ 0.723   $ & $ -0.235  $ & $  -0.279  $ & $  -0.272  $ & $  -0.272  $ \\ 
$ p_{70}  $ & $  37.7  $ & $  15.3  $ & $ 22.1   $ & $ 42.7  $ & $  18.6  $ & $  33.8  $ & $  21.2  $ \\ 
$ p_{80} $ & $  7.69  $ & $  7.23  $ & $ 8.53   $ & $ 12.0  $ & $  4.23  $ & $  10.1  $ & $  4.07  $ \\ 
\end{tabular}
\end{ruledtabular}
\end{table*}

\begin{table*}
\caption{\label{table:num:dist} Numerical values of the energy and rapidity distribution fit parameters $p_{ij}$ (refitted modified energy spectrum)
and $q_{ij}$. A hyphen indicates that the parameter is not used in the parameterization.}
\begin{ruledtabular}
\begin{tabular}{ c  c c c  c c c c }
 &$\pi^+ $ & $ \pi^-$ & $\pi^0$ &  $K^+ $& $K^-$ & $K^0$ & $\overline{K^0} $  \\
\hline
$ p_{00} $ & $ -0.474 $ & $ -0.461 $ & $ -0.420 $ & $ -1.03 $ & $ -1.20 $ & $ -1.13 $ & $ -1.37 $ \\ 
$ p_{01} $ & $ 0.0846 $ & $ 0.0796 $ & $ 0.0821 $ & $ -0.00299 $ & $ 0.0377 $ & $ 0.0168 $ & $ 0.0501 $ \\ 
$ p_{10} $ & $ -0.115 $ & $ -0.124 $ & $ -0.118 $ & $ -0.0375 $ & $ -0.0291 $ & $ -0.0604 $ & $ -0.0579 $ \\ 
$ p_{11} $ & $ 0.0102 $ & $ 0.0117 $ & $ 0.0107 $ & $ 0.00396 $ & $ -0.000110 $ & $ 0.00621 $ & $ 0.00351 $ \\ 
$ p_{20} $ & $ -0.560 $ & $ -0.604 $ & $ -0.598 $ & $ -0.835 $ & $ -0.606 $ & $ -0.655 $ & $ -0.467 $ \\ 
$ p_{21} $ & $ 0.497 $ & $ 0.496 $ & $ 0.497 $ & $ 0.494 $ & $ 0.497 $ & $ 0.497 $ & $ 0.497 $ \\ 
$ p_{30} $ & $ -1.15 $ & $ -0.641 $ & $ -0.815 $ & $ -0.742 $ & $ -0.845 $ & $ -0.788 $ & $ -0.917 $ \\ 
$ p_{40} $ & $ -1.03 $ & $ -1.11 $ & $ -1.17 $ & $ -0.167 $ & $ -0.155 $ & $ -0.237 $ & $ -0.176 $ \\ 
$ p_{50} $ & $ 1.12 $ & $ 0.980 $ & $ 0.987 $ & $ 0.716 $ & $ 0.934 $ & $ 0.840 $ & $ 1.01 $ \\ 
$ p_{60} $ & $ 0.962 $ & $ 0.891 $ & $ 0.954 $ & $ 1.08 $ & $ 0.597 $ & $ 0.789 $ & $ 0.371 $ \\
$ p_{70}$ & - & - & - & - & - & - & -   \\
$ p_{80}$ & 6.98 & 6.93 & 7.45 & - & - & - & -   \\
\hline
$ q_{00} $ & $ -0.167 $ & $ -0.149 $ & $ -0.161 $ & $ 0.539 $ & $ 0.363 $ & $ 0.405 $ & $ 0.382 $ \\ 
$ q_{01} $ & $ 0.0497 $ & $ 0.108 $ & $ 0.0737 $ & $ 0.222 $ & $ 0.228 $ & $ 0.149 $ & $ 0.195 $ \\ 
$ q_{10} $ & $ 0.668 $ & $ 0.668 $ & $ 0.637 $ & $ 0.889 $ & $ 1.06 $ & $ 0.997 $ & $ 1.12 $ \\ 
$ q_{11} $ & $ 0.329 $ & $ 0.328 $ & $ 0.307 $ & $ 0.523 $ & $ 0.673 $ & $ 0.612 $ & $ 0.719 $ \\ 
$ q_{12} $ & $ 0.116 $ & $ 0.0806 $ & $ 0.107 $ & $ 0.227 $ & $ 0.328 $ & $ 0.216 $ & $ 0.298 $ \\ 
$ q_{13} $ & $ -0.162 $ & $ -0.154 $ & $ -0.144 $ & $ -0.304 $ & $ -0.141 $ & $ -0.306 $ & $ -0.155 $ \\ 
$ q_{20} $ & $ 2.03 $ & $ 2.05 $ & $ 2.16 $ & $ 0.902 $ & $ 0.695 $ & $ 0.676 $ & $ 0.579 $ \\ 
$ q_{21} $ & $ -0.0577 $ & $ -0.0654 $ & $ -0.0704 $ & $ -0.0694 $ & $ -0.0648 $ & $ -0.0525 $ & $ -0.0527 $ \\ 
$ q_{22} $ & $ 0.247 $ & $ 0.233 $ & $ 0.216 $ & $ 0.185 $ & $ 0.226 $ & $ 0.242 $ & $ 0.261 $ \\ 
$ q_{23} $ & $ 0.665 $ & $ 0.381 $ & $ 0.556 $ & $ 0 $ & $ 0 $ & $ 0 $ & $ 0 $ \\ 
$ q_{30} $ & $ 1.04 $ & $ 1.24 $ & $ 1.37 $ & $ 0.319 $ & $ 0.198 $ & $ 0.238 $ & $ 0.184 $ \\ 
$ q_{31} $ & $ 3.94 $ & $ 4.51 $ & $ 4.92 $ & $ 1.16 $ & $ 1.17 $ & $ 0.951 $ & $ 1.08 $ \\ 
$ q_{32} $ & $ -1.37 $ & $ -1.54 $ & $ -1.65 $ & $ -0.597 $ & $ -0.699 $ & $ -0.559 $ & $ -0.684 $ \\ 
\end{tabular}
\end{ruledtabular}
\end{table*}

\end{document}